\documentclass[useAMS,usenatbib]{mn2e}

\usepackage{aas_macros}
\usepackage{amsmath}
\usepackage{amssymb}
\usepackage{epsfig}
\usepackage{graphicx}
\usepackage{ifthen}
\usepackage{latexsym}
\usepackage{ogonek}
\usepackage{rotating}
\usepackage{subfigure}
\usepackage{times,epsf}
\usepackage{txfonts}
\usepackage{varioref}
\usepackage{verbatim}
\usepackage{url}
\newcommand{\res}{$n_r\times n_\theta \times n_\phi$}
\newcommand{\rgammie}{r_B}

\newcommand{\shear}{\sigma_{\hat{r}\hat{\phi}}}

\newcommand{\htt}{{\hat{t}}}
\newcommand{\hr}{{\hat{r}}}
\newcommand{\hh}{{\hat{\theta}}}
\newcommand{\hp}{{\hat{\phi}}}

\newcommand{\tacc}{t_{\rm visc}}
\newcommand{\tchunk}{t_{\rm chunk}}
\newcommand{\rloose}{r_{\rm loose}}
\newcommand{\rsevere}{r_{\rm strict}}

\newcommand{\faraday}{F^{\mu\nu}}

\newcommand{\thin}{A}
\newcommand{\thinA}{B}
\newcommand{\thinLR}{C}
\newcommand{\sane}{D}
\newcommand{\saneA}{E}
\newcommand{\mad}{F}

\newcommand{\trey}{T^{\rm (rey)}}
\newcommand{\tmax}{T^{\rm (mag)}}

\begin{document}

\title[Shakura-Sunyaev viscosity prescription] {The Shakura-Sunyaev
viscosity prescription with variable $\alpha(r)$} 
\author[R.~F. Penna, A.~S. S\k{a}dowski, A.~K. Kulkarni, \& R. Narayan] 
{Robert F. Penna$^1$ 
\thanks{E-mail: rpenna@cfa.harvard.edu~(RFP),
asadowski@cfa.harvard.edu~(AS), narayan@cfa.harvard.edu~(RN),},
Aleksander S\k{a}dowski$^1$\footnotemark[1], 
Akshay K. Kulkarni, 
Ramesh Narayan$^1$\footnotemark[1]\\
$^1$Harvard-Smithsonian Center for Astrophysics, 60 Garden Street,
Cambridge, MA 02138, USA \\ } 

\date{\today}

\maketitle

\begin{abstract}

Almost all hydrodynamic accretion disk models parametrize viscosity
with the dimensionless parameter $\alpha$. There is no detailed model
for $\alpha$, so it is usually taken to be a constant.  However, global
simulations of magnetohydrodynamic disks find that $\alpha$ varies
with distance from the central object.  Also, Newtonian simulations
tend to find smaller $\alpha$'s than general relativistic simulations.
We seek a one-dimensional model for $\alpha$ that can reproduce these
two observations.
We are guided by data from six general relativistic
magnetohydrodynamic accretion disk simulations.  The variation of
$\alpha$ in the inner, laminar regions of the flow results from
stretching of mean magnetic field lines by the flow.  The variation of
$\alpha$ in the outer, turbulent regions results from the dependence
of the magnetorotational instability on the dimensionless shear rate.
We give a one-dimensional prescription for $\alpha(r)$ that captures
these two effects and reproduces the radial variation of $\alpha$
observed in the simulations.  For thin disks, the prescription
simplifies to the formula $\alpha(r)=0.025 \left[q(r)/1.5\right]^6$,
where the shear parameter, $q(r)$, is an analytical function of radius
in the Kerr metric.  The coefficient and exponent are inferred from
our simulations and will change as better simulation data becomes
available.
We conclude that the $\alpha$-viscosity prescription can be extended
to the radially varying $\alpha$'s observed in simulations.  It is
possible that Newtonian simulations find smaller $\alpha$'s than
general relativistic simulations because the shear parameter is lower
in Newtonian flows.
\end{abstract}


\section{Introduction}
\label{sec:intro}

Accretion disk magnetic fields and turbulence act as a large scale
viscosity, draining angular momentum and energy from accreting gas.
The magnitude of this viscosity is uncertain, which adds a free
parameter to accretion disk models.  For example,
\citet{pringlerees1972} leave the ratio of the disk's radial velocity
to its circular velocity as a free parameter.  They call this ratio
$y/100$ and estimate $y\sim 1$.  They note that $y$ might be radius
dependent and that a strong radial dependence could change qualitative
features of the disk.  However, other important features of the disk,
such as its luminosity, depend weakly or not at all on $y$, so
progress is possible without a detailed model for $y$.

\citet{SS73} parametrized the viscosity with $\alpha$, the ratio of
stress to pressure.  It is related to Pringle and Rees's $y$ by $y=100
\alpha (h/r)^2$, where $h/r$ is the disk opening angle.
\footnote{\citet{pringlerees1972} leave $h/r$ a free parameter which
they suppose to be $\sim 0.05$.  \citet{SS73} use hydrostatic
equilibrium to solve for $h/r$ self-consistently.}  They
anticipated that $\alpha$ will be a function of radius, citing
experiments of turbulence in Taylor-Coutte flow, flow between rotating
cylinders \citep{taylor36}.  These experiments show that the torque
exerted by a turbulent flow on the cylinders depends on the rotation
rate of the cylinders and the separation between them.  The torque in
the experiment is related to $\alpha$, so $\alpha$ should depend on
accretion flow properties that vary with radius.  The analogy is too
rough to produce a quantitative model, so they assume $\alpha$ is a
constant for simplicity.

The most significant theoretical breakthrough towards an understanding
of $\alpha$ has been the realization that the magnetorotational
instability (MRI) can drive turbulence in ionized accretion disks
\citep{velikhov1959,chandrasekhar1960,balbus1991,balbus1998}.  A weak seed magnetic
field and a radially decreasing angular velocity profile are all that
are required to trigger the instability and both are present in disks.
This makes it possible to run magnetohydrodynamic (MHD) disk
simulations without using the $\alpha$-viscosity prescription.

Nonetheless, because of its relative simplicity, the
$\alpha$-viscosity prescription has not dimmed in importance.  Over
the past year, more than 300 papers cited the pioneering work of
\citet{SS73} and more than 1000 papers made reference to some disk
solution based on the $\alpha$-viscosity prescription (such as the
relativistic thin disk of \citet{nt73} or the advection dominated
accretion flow (ADAF) of \citet{ny1994}).  However, despite its
interest, there is still no widely accepted model for the size or
shape of $\alpha$.  In applications, it is typically assumed to be a
constant between 0.01 and 0.1 \citep[e.g.,][]{gou2011}.

%5
There are two varieties of MHD simulations, local and global, and both
provide hints about the size and shape of $\alpha$.  A standard setup
for a local simulation is a box of weakly magnetized fluid in the
shearing sheet approximation
\citep[e.g.,][]{hb1991,brandenburg1995,hawley1995,hawley1996,stone1996,brandenburg2001,sano2004}.
The dimensions of the box are typically a few disk scale heights. The
earliest global simulations of MRI turbulent disks used a Newtonian
potential \citep{armitage1998,matsumoto1999,hawley2000,machida2000}
and, later, a pseudo-Newtonian potential \citep{hawley2001}.  The
first general relativistic global simulations were carried out by
\citet{dvhk2003}.  A standard setup for global simulations is a torus
of fluid in hydrostatic equilibrium, threaded with a weak magnetic
field \citep[e.g.,][]{dvhk2003}.  In local and global simulations,
differential rotation triggers the MRI and drives turbulence.  When
the simulation reaches a quasi-steady state, one can compute the ratio
of stress to pressure and so measure $\alpha$.  Local simulations
focus on resolving small scale physics, such as saturation of the MRI,
while global simulations attempt a complete portrait of the accretion
disk.

%M1
Global simulations have hinted at the shape of $\alpha$.
\citet{penna2010} and \citet{penna2012} presented general relativistic
MHD (GRMHD) simulations of MRI turbulent disks in the Kerr metric, the
spacetime of a spinning black hole.  They confirmed that the
relativistic $\alpha$-disk model of \citet{nt73} gives a good
description of the simulation data.  This was partly motivated by
earlier suggestions that magnetic stresses at the inner edges of MHD
disks invalidate the assumptions of $\alpha$-disk models 
\citep[e.g.,][]{krolik1999,agol2000}.  The GRMHD simulations produced
luminosity inside the innermost stable circular orbit (ISCO), where
the \citet{nt73} model is entirely dark, but it was a modest
contribution to the total luminosity of the disk \citep[on the order of a
few percent;][]{kulkarni2011}.  Later, \citet{penna2012} generalized the \citet{nt73}
model to self-consistently incorporate nonzero luminosity inside the
ISCO, bringing it closer to the GRMHD simulations.

%M2
\citet{penna2010} and \citet{penna2012} measured the shape of
$\alpha$.  For accretion onto a non-spinning black hole, they found
$\alpha$ is small at the event horizon, increases to a maximum near the
photon orbit, declines to $\sim 0.06$ near the ISCO, and then
continues to decline, albeit more slowly \citep[see, e.g., Figure 7 of
][] {penna2012}.  In a completely different context, MHD simulations
of protoplanetary disks, \citet{fromang2011} found a radially varying
$\alpha$ with similar shape.  However, the overall size of their
$\alpha$ was over an order of magnitude lower, peaking at 0.013 and
declining to below 0.002.  

%M3
The value of $\alpha$ is notoriously difficult to pin down
\citep{pessah2007}.  Local simulations find that $\alpha$ is strongly
affected by grid scale dissipation as well as stratification
\citep{lesur2007,fromang2007,simon2009,davis2010}.  In global and
local simulations, $\alpha$ depends non-monotonically on resolution at
all but the highest resolutions \citep{sorathia2012}.  In models with
net magnetic flux, $\alpha$ scales with increasing flux
\citep{hawley1995,sano2004,pessah2007}.  Finally, $\alpha$ depends on
the initial magnetic field geometry and strength
\citep[e.g.,][]{sorathia2012}.

%M4
A combination of these effects can probably explain some of the
discrepancy between the $\alpha$ values found by \citet{penna2012} and
\citet{fromang2011} . The former employed an initially poloidal
magnetic field with initial gas-to-magnetic pressure ratio $\beta=100$
and resolution $256\times 64\times 32$ in $(r,\theta,\phi)$.  The
later employed an initially toroidal field with initial $\beta=25$ and
resolution $512\times 256\times 256$.  Neither had explicit
dissipation or net magnetic flux.  Initially toroidal fields tend to
beget smaller $\alpha$'s than initially poloidal fields, but $\alpha$
also tends to be inversely related to the initial $\beta$.  The
scaling of $\alpha$ with resolution is non-monotonic
\citep{sorathia2012}.  It is not clear whether these effects are large
enough to explain why the $\alpha$'s measured from the general
relativistic simulations are over an order of magnitude larger than
the $\alpha$'s measured from the Newtonian simulations.

%M5
In this paper, we present a one-dimensional model for the shape of
$\alpha$ and we show that relativistic corrections enhance the
$\alpha$'s measured in GRMHD simulations relative to Newtonian
simulations.  The one-dimensional model has two components.  The first
component is generated by large-scale, mean magnetic fields and is
based on the model of \citet{gammie1999}.  It dominates in the inner
regions of accretion flows, where plunging gas stretches and amplifies
the frozen-in magnetic field.  The second, ``turbulent,'' component
describes the dependence of $\alpha$ on the shear rate of the flow.
The dependence of $\alpha$ on the shear rate in turbulent flows was
earlier emphasized by \citet{godon1995}, \citet{abramowicz1996}, and \citet{pessah08}.
Combining the mean field and turbulent components yields a
one-dimensional model for the shape of $\alpha$.

%M6
We use this model to fit the profiles of $\alpha$ versus radius
extracted from six GRMHD simulations.  The simulations describe thin
disks accreting onto a non-spinning black hole at two different
resolutions, a thin disk and a thick disk accreting onto a spinning
black hole (with dimensionless spin parameter $a/M=0.7$), and thick
disks with two different initial magnetic field topologies accreting
onto a nonspinning black hole.  We show how the radial variation of
$\alpha$ changes the structure of $\alpha$-disk solutions.  Finally,
we note that the enhancement of $\alpha$ seen in GRMHD simulations
relative to Newtonian simulations can be explained by
the dependence of $\alpha$ on shear rate.

%M7
There is another reason $\alpha$ is interesting which we have not yet
mentioned.  Turbulent fluids can be complicated.  They are disordered
solutions of nonlinear equations that require great effort to solve
numerically.  And yet, from fully developed turbulence, simple scaling
laws emerge with apparently universal properties.  For example,
shearing box simulations of MRI-driven turbulence with a Keplerian
rotation profile always find that the ratio of Maxwell stress to
Reynolds stress is a constant, $\approx 4$ \citep{pessah2006}.  The
same ratio appears independently of the magnitude or geometry of the
magnetic field.  It seems only to depend (in a simple way) on the
shear rate of the flow \citep{hawley1999,pessah2006}.  Similarly, the
viscosity parameter, $\alpha$, obeys a remarkable scaling law: $\alpha
\beta \approx 1/2$, where $\beta$ is the gas-to-magnetic pressure
ratio \citep{blackman2008,guan2009,sorathia2012}.  Formulas this
simple should have simple explanations; they should not just appear at
the end of large numerical calculations, as if by coincidence.  We
hope that clarifying some of the physics underlying $\alpha$ will
improve our understanding of these scaling laws.

The paper is organized as follows.  In \S\ref{sec:prelims}, we give an
overview of physics in the Kerr metric.  In \S\ref{sec:1dalpha}, we
describe our one-dimensional model for $\alpha(r)$.  The GRMHD
simulations are described in \S\S\ref{sec:thesims}-\ref{sec:moresims};
we give a broad overview of the simulations in \S\ref{sec:thesims},
analyze the two fiducial simulations in \S\ref{sec:fidsims}, and
analyze the remaining four simulations in \S\ref{sec:moresims}.  In
\S\ref{sec:slim}, we show how a radially varying $\alpha$ affects the
structure of $\alpha$-disk solutions.  We conclude with a summary and
discussion in \S\ref{sec:discuss}.

\section{Preliminaries}
\label{sec:prelims}

%T1
For our investigation of $\alpha$, we will need to compute the angular velocity,
shear rate, and epicyclic frequency of an accreting gas in the Kerr
metric, so we review their definition here.  This also helps to
establish notation.  As an intermediate step, we discuss the
transformation between the Boyer-Lindquist and fluid frames.

The Kerr metric in Boyer-Lindquist coordinates is
\begin{align}\label{eq:kerr}
ds^2 &= -\left(1-2Mr/\Sigma \right) dt^2
       -\left(4Mar \sin^2\theta/\Sigma\right)dtd\phi
       +\left(\Sigma/\Delta\right)dr^2 \notag \\
      &+\left(r^2+a^2+2Ma^2r\sin^2\theta/\Sigma\right)\sin^2\theta d\phi^2.
\end{align}
Here $M$ is the mass of the black hole, $a$ is its angular momentum per
unit mass $(0\leq a \leq M)$, and the functions $\Delta,\, \Sigma$, and  $A$
are defined by
\begin{align}
\Delta &\equiv r^2 - 2Mr +a^2,\label{eq:delta}\\
\Sigma &\equiv r^2+a^2\cos^2\theta,\\
A &\equiv \left(r^2+a^2\right)^2-a^2\Delta \sin^2 \theta.\label{eq:A}
\end{align}
The accreting, magnetized gas is characterized by its four-velocity,
$u^\mu$, density, $\rho$, pressure, $p$, and internal energy, $u$,
and by the electromagnetic field, $F^{\mu\nu}$.  We set $G=c=1$.  

%T2
The angular velocity of the gas is
\begin{equation}\label{eq:omega}
\Omega \equiv \frac{d\phi}{dt} = \frac{u^\phi}{u^t}.
\end{equation}
Circular equatorial geodesics have
$\Omega=M^{1/2}/\left(r^{3/2}+aM^{1/2}\right)$, and the motion of a
geometrically thin disk is well approximated by circular geodesics
outside the ISCO.  However, motion inside the ISCO and the motion
of thick disks are not Keplerian, so we leave $\Omega$ unspecified for
now.

\subsection{Inertial Fluid Frame}

%T3
We would like to evaluate the shear rate and epicyclic frequency in
the inertial fluid frame, rather than the Boyer-Lindquist, ZAMO, or
any other frame, because the physics is simplest there.  Also, this
is the frame where $\alpha$ is defined.  In this frame, the equivalence
principle lets us ignore gravitational forces at the center of the
fluid parcel we are following.  Tidal forces and other gravitational
forces that become important over large distances will not concern us
because shear rate and epicyclic frequency are local measurements.

%T4
Measurements in the Boyer-Lindquist frame,
$\left(dt,dr,d\theta,d\phi\right)$, are related to measurements in the
inertial fluid frame, $\left(\omega^\htt,\omega^\hr,\omega^\hh,\omega^\hp\right)$, by the
the transformation matrix $\omega^\mu_{\hat{\mu}}$
\citep{krolik2005,beckwith2008,kulkarni2011}:
\begin{align}
\omega^\mu_{\hat{t}} &= \left(u^t,u^r,u^\theta,u^\phi\right), \label{eq:et}\\
\omega^\mu_{\hat{r}} &= \frac{s}{N_1}
       \left(u_r u^t,
       1+u^r u_r+u^\theta u_\theta,
       0,
       u_r u^\phi\right), \label{eq:er}\\
\omega^\mu_{\hat{\theta}} &= \frac{1}{N_2}
       \left(u_{\theta}u^t,u_\theta u^r,
       1+u_\theta u^\theta,
       u_\theta u^\phi\right),\label{eq:eh} \\
\omega^\mu_{\hat{\phi}} &= \frac{1}{N_3}\left(-\ell,0,0,1\right),\label{eq:ep}
\end{align}
where,
\begin{align}
s &= -C_0/\left|C_0\right|,& \ell &= u_\phi/u_t, \notag \\ 
N_1 &= g_{rr}\sqrt{g_{tt}C_1^2+g_{rr}C_0^2
          +g_{\phi\phi}C_2^2+2g_{t\phi}C_1C_2},&
C_0 &= u^t u_t+u^\phi u_\phi, \notag\\
N_2 &= \sqrt{g_{\theta\theta}\left(1+u^\theta u_\theta\right)}, &
C_1 &= u^r u_t, \notag \\
N_3 &= \sqrt{g_{tt} \ell^2 - 2g_{t\phi}\ell -g_{\phi\phi}},
           &  C_2 &= u^r u_\phi.
\end{align}
Hatted indices refer to fluid frame quantities and unhatted indices
refer to Boyer-Lindquist frame quantities.  In the orthonormal fluid
frame the metric is the Minkowski metric,
$\eta_{\hat{a}\hat{b}}={\rm diag}(-1,1,1,1)$.  So hatted indices
are raised and lowered with the Minkowski metric and unhatted indices
are raised and lowered with the Kerr metric.

%T5
The fluid frame basis \eqref{eq:et}-\eqref{eq:ep} was constructed
using a Gram-Schmidt process.  There is some arbitrariness in the
orientation of the frame, but we followed standard conventions.
Equation \eqref{eq:et} for $\omega_\htt$ is necessary because the Lorentz
factor in the fluid frame should satisfy $\gamma=-u_\htt=-\omega^\mu_\htt
u_\mu = 1$.  The next step in the Gram-Schmidt process is to define
$\omega_\hp$ such that it is orthogonal to $\omega_\htt$ and has no component
along $dr$ or $d\theta$.  Finally, $\omega_\hr$ and $\omega_\hh$ are
constructed.  Notice that $\omega_\hr$ has a nonzero component along
$d\phi$, and $\omega_\hh$ has components along all four Boyer-Lindquist
directions.  This is unavoidable.  However, the most important
directions for the physics, $\omega_\hr$ and $\omega_\hp$, are aligned as
closely as possible with their Boyer-Lindquist analogues.

%T6
The arbitrariness in the construction of the fluid frame
leads to an ambiguity in the definition of $\alpha$.  One usually
avoids quantities with these sorts of ambiguities.  But, as
discussed in \S\ref{sec:intro}, $\alpha$ is too useful for accretion
disk modeling and turbulence theory to abandon.  So our strategy
is to define $\alpha$ in the most natural way possible and see
where this leads.

%T7
If the poloidal velocity is much smaller than the azimuthal velocity,
as usually happens everywhere except in the innermost regions of the
disk, the fluid frame basis simplifies \citep{nt73}:
\begin{align}
\omega^\htt &= u_t dt +u_\phi d\phi, \label{eq:et2}\\
\omega^\hr &= \mathcal{D}^{-1/2}dr, \\
\omega^\hh &= r \, d\theta, \\
\omega^\hp &= \gamma r \mathcal{A}^{1/2}
          \left(d\phi -\Omega dt\right).\label{eq:ep2}
\end{align}
The Lorentz factor is $\gamma=\sqrt{-g_{tt}}u^t$ and
\begin{equation}
\mathcal{A} = 1+a^2r^{-2}+2Ma^2r^{-3}, \quad
\mathcal{D} = 1-2Mr^{-1}+a^2r^{-2}.
\end{equation}
The relativistic factors $\mathcal{A}, \mathcal{D}$, and $\gamma$ are
unity at large radii.  In fact, in this limit, the
fluid frame basis is exactly aligned with the Boyer-Lindquist frame.
So the ambiguities in the general relativistic definition of $\alpha$
discussed earlier are not important in the outer regions of the disk.
There remains a potential ambiguity in the construction of the Boyer-Lindquist
$r$ and $\theta$ coordinates, even when the gas is nonrelativistic,
because the black hole's spin axis breaks the spherical symmetry of
spacetime. But this is also negligible far from the black hole where
frame dragging is weak.

\subsection{Electric and Magnetic Fields}
\label{sec:EM}

%R3
All observers interact with the same electromagnetic field.  They
might measure its components, $F_{\mu\nu}$, differently, depending on
their reference frames, but the underlying object is the same
multilinear map, $\mathbb{F}$.

%R4
The electric and magnetic fields are not so universal: each observer
splits the electromagnetic field into different electric and magnetic
components.  An observer with four-velocity $u^\mu$ measures electric
and magnetic fields
\begin{equation}\label{eq:eb}
e^\mu = u_\nu F^{\nu\mu}, \quad b^\mu = u_\nu  \phantom{}_*\faraday,
\end{equation}
where the dual Faraday tensor is
$\phantom{}_* F^{\mu\nu}=\epsilon^{\mu\nu\kappa\lambda}F_{\kappa\lambda}/2$, the
Levi-Civita symbol is
$\epsilon^{\mu\nu\kappa\lambda}=-\left[\mu\nu\kappa\lambda\right]/\sqrt{-g}$,
and $\left[\mu\nu\kappa\lambda\right]$ is the completely antisymmetric
symbol, equal to either $0, -1$, or $+1$.  

The electric and magnetic fields $e^\mu$ and $b^\mu$ transform as
tensors, so they can be evaluated in any frame. In general, all four
components are nonzero.  The four-velocity of the observer, $u^\mu$,
appears in \eqref{eq:eb}, so each observer interacts with different
electric and magnetic fields.  Following standard convention, we denote
the Boyer-Lindquist observer's splitting $E^\mu$ and $B^\mu$, and we
denote the fluid frame's splitting $e^\mu$ and $b^\mu$.  In
Boyer-Lindquist coordinates $B^t=E^t=0$, and in the fluid frame
$e^\htt=b^\htt=0$, by the antisymmetry of $F^{\mu\nu}$.  The remaining
nonzero components in these frames are the usual three-vectors of
special relativity.

%R6
In ideal MHD, $e^\mu=0$, so the fluid frame splitting of $F_{\mu\nu}$
is usually the simplest.  For calculations, Boyer-Lindquist
coordinates are sometimes more convenient than the fluid frame.  So it
is common to work with $b^\mu$, the fluid frame observer's magnetic
field expressed in Boyer-Lindquist coordinates.  This is legitimate,
because $b^\mu$ is a 4-vector, but no observer would measure $b^t,
b^r, b^\theta$, or $b^\phi$.  Physically meaningful quantities are
$b^\hr, b^\hh$, and $b^\hp$.

%R7
It is possible to convert between $B^\mu$ and $b^\mu$ directly, without
reference to $F^{\mu\nu}$.  In Boyer-Lindquist coordinates, the
formulae are:
\begin{align}
b^t   &= B^\mu u^\nu g_{\mu \nu}, \quad 
b^i = \frac{B^i + b^t u^i}{u^t}, \label{eq:Btob}\\
B^\mu &= b^\mu u^t - b^t u^\mu.\label{eq:btoB}
\end{align}
In these equations, $i$ runs over $r,\theta,\phi$, and $\mu, \nu$ run
over $t, r, \theta, \phi$.

\subsection{Shear Rate}
\label{sec:shear}

%T8
The shear rate is a measure of how rapidly the angular velocity of an
accretion disk varies with radius.  It is $r\Omega_{,r}/2$ in
Newtonian gravity.  More generally, one can define the special
relativistic shear tensor
\begin{equation}
\sigma_{\alpha\beta} \equiv \frac{1}{2}\left(u_{\alpha,\mu}h^\mu_\beta 
+u_{\beta,\mu}h^\mu_\alpha\right)-\frac{1}{3}\Theta h_{\alpha\beta},\label{eq:sigma}
\end{equation}
where $h_{\alpha\beta}=g_{\alpha\beta}+u_\alpha u_\beta$ is the
projection tensor and $\Theta={u^\alpha}_{;\alpha}$ is the expansion
scalar \citep{nt73}.  Then the shear rate is the $\hr\hp$ component of
this tensor measured in the fluid frame: $\shear$.  The shear
tensor so defined is trace-free and symmetric.  To obtain the general
relativistic version, one should replace partial derivatives in
equation \eqref{eq:sigma} with covariant derivatives.  We do not
require this generality because in the inertial fluid frame the laws
of physics take on their special relativistic form without gravity, by
the equivalence principle.

%T9
To obtain a simple formula for the shear rate, let us assume the
poloidal velocity is small and the flow is axisymmetric.  Then we can
ignore the expansion scalar, $\Theta$, use equations
\eqref{eq:et2}-\eqref{eq:ep2} for the fluid frame transformation, and ignore
derivatives with respect to $\hat{\phi}$.  The fluid frame projection
tensor is $h_{\hat{\alpha}\hat{\beta}}=\rm{diag}\left(0,1,1,1\right)$,
so the shear rate is
\begin{equation}
\shear =\frac{1}{2} {u^\hp}_{,\hr},
\end{equation}
which resembles the Newtonian shear rate, $r\Omega_{,r}/2$.  In the
fluid frame, $u^\hp (r)=0$, so the derivative is
\footnote{We abuse notation for clarity.  The commutators of fluid
frame basis elements do not vanish in general, so they are not
coordinate induced: there is no ``$\hat{r}$'' coordinate satisfying
$d\hat{r}=e^{\hat{r}}$.}:
\begin{equation}
%A  =\frac{1}{2}\frac{u^{\hat{\phi}}(r+d\hat{r})-u^{\hat{\phi}}(r)}{d\hat{r}}
\shear  = \frac{1}{2}\lim_{d\hat{r}\to 0}\frac{u^{\hat{\phi}}(r+d\hat{r})}{d\hat{r}}.
\end{equation}
We use equations \eqref{eq:et2}-\eqref{eq:ep2} to rewrite
fluid frame measurements in terms of Boyer-Lindquist measurements,
obtaining:
\begin{equation}
%A  = \frac{1}{2}\gamma r \sqrt{\mathcal{A}}
%       \frac{u^\phi(r+d\hat{r})-\Omega(r) u^t(r+d\hat{r})}{d\hat{r}}
%A  = \frac{1}{2}\gamma^2 r \mathcal{A}\frac{\Omega(r+dr)-\Omega(r)}{dr}
\shear = \frac{1}{2}\gamma^2\mathcal{A} r \thinspace\Omega_{,r}
.\label{eq:shearrate}
\end{equation}
This is the product of the Newtonian shear rate $r\Omega_{,r}/2$ with
a relativistic correction, $\gamma^2 \mathcal{A}$.  \citet{nt73} state
this formula without proof.  It describes the rate of change of a
disk's angular velocity with radius, assuming the azimuthal velocity
dominates the poloidal velocity.

%W1
A dimensionless measure of the shear rate is
\begin{equation}\label{eq:q}
q =  - 2\shear/\Omega  = - \gamma^2 \mathcal{A}
\frac{d\log\Omega}{d\log r}.
\end{equation}
Positive $q$ corresponds to angular velocity decreasing with
radius. Solid body rotation is $q=0$.  Accretion disks generally have
positive $q$, although $q$ changes sign near the photon orbit in black
hole disks.  Flows with positive $q$ are unstable to the MRI
\citep{velikhov1959,chandrasekhar1960,balbus1991}, and the value of
$\alpha$ is a function of $q$ \citep{pessah08}.  Flows with $q>2$ are
Rayleigh (hydrodynamically) unstable.  The MRI has been analyzed in
the $q>2$ regime by \citet{balbus2012}\footnote{Note that our
simulated disks have $q<2$ at all radii.  The scaled epicyclic
frequency reaches a minimum near the ISCO and increases in the
plunging region.}.  Circular geodesics in Newtonian gravity have
$q=3/2$.  Circular, equatorial geodesics in the Kerr metric have
\citep{gammie04}
\begin{equation}\label{eq:qcirc}
q=\frac{3}{2}\frac{1-2Mr^{-1}+a^2r^{-2}}{1-3Mr^{-1}+2aM^{1/2}r^{-3/2}}.
\end{equation}
which becomes $3/2$ at large radii.  The fact that $\alpha$ is a
function of $q$, and in the Kerr metric $q$ is a function of radius,
is the basis for the turbulent component of the one-dimensional
$\alpha$ model in \S\ref{sec:1dalpha}.

\subsection{Radial Epicyclic Frequency}

%W2
The radial epicyclic frequency is the frequency at which a radially displaced
fluid parcel will oscillate. 
As a function of $q$, it is \citep{pringle2007}
\begin{equation}\label{eq:epicyclic}
\kappa=\sqrt{2(2-q)} \Omega.
\end{equation}
This expression is valid in the inertial fluid frame
provided we use the relativistic expressions for 
$\Omega$ and $q$, equations \eqref{eq:omega} and \eqref{eq:q}.  

Circular geodesics at the ISCO have $\kappa=0$ by definition, which
makes it easy to see that they have $q=2$.  A thin accretion disk
starts with $q=3/2$ at large radii and then $q$ increases until it
reaches $2$ at the inner edge.  The radial dependence of $\kappa^2$
for circular, equatorial Kerr geodesics is \citep{gammie04}
\begin{equation}
\kappa^2=\frac{1}{r^3}
\frac{1-6M/r+8aM^{1/2}r^{-3/2}-3a^2/r^2}
{1-3Mr^{-1}+2aM^{1/2}r^{-3/2}}.
\end{equation}
The epicyclic frequency is zero at the ISCO, and imaginary inside
the ISCO, signaling the instability of circular geodesics.  In
\S\ref{sec:fidsims} and \S\ref{sec:moresims}, we compute the epicyclic
frequency of simulated GRMHD accretion disks as a function of radius.
The inner edge of the disk can be identified with the minimum
$\kappa$.  Only for thin disks does $\kappa$ have a sharp minimum at
the ISCO.  In general, pressure gradient forces and magnetic stresses
smear out the inner edge of the disk and displace it from the
ISCO.

\section{One-Dimensional Model For $\alpha(\lowercase{r})$}
\label{sec:1dalpha}

%W4
In this section we define a one-dimensional prescription for the
dependence of $\alpha$ on radius.  Our model is the sum of two
components: a turbulent component that dominates in the outer regions
of the disk and a large scale magnetic field component that dominates
in the inner regions of the disk.  We discuss each component
separately before combining them into a single prescription for
$\alpha(r)$.  We will fit this $\alpha(r)$ prescription to data
from GRMHD simulations in \S\S\ref{sec:thesims}-\ref{sec:moresims}.

The standard $\alpha$ viscosity prescription is
\begin{equation}\label{eq:SSalpha}
T_{\hat{r}\hat{\phi}} = \alpha p,
\end{equation}
where $T_{\hat{r}\hat{\phi}}$ is the fluid frame stress, $p$ is
pressure, and $\alpha$ is a constant \citep{SS73}.  There are two
equivalent ways to modify this prescription.  One can add extra
factors multiplying the RHS of equation \eqref{eq:SSalpha} and keep
$\alpha$ a constant, or one can keep equation \eqref{eq:SSalpha}
unchanged but define $\alpha$ to be a function of radius,
$\alpha=\alpha(r)$.  We chose the later convention because it makes
it easy to adapt old $\alpha$-disk solutions to the new prescription:
just insert $\alpha(r)$ wherever $\alpha$ appears.  Both approaches
have been used in the past so it requires a bit of care to compare
results.  For example, \citet{abramowicz1996} discussed the dependence
of $\alpha$ on $q$, while \citet{pessah08} modified \eqref{eq:SSalpha}
and kept $\alpha$ a constant.

\subsection{Turbulent $\alpha$}
\label{sec:turbalpha}

%W5
We have seen in \S\ref{sec:shear} that $q$ can be a function of
radius, either through relativistic corrections to the Newtonian shear
rate, or through the dependence of $\Omega$ on radius.  It turns out
$\alpha$ is a function of $q$, and so it too can depend on radius.
That $\alpha$ depends on $q$ is perhaps not surprising.  MHD flows are
unstable to the MRI and have finite $\alpha$ when $q>0$, but they are
stable against the MRI and have vanishing $\alpha$ when $q<0$.  Some
dependence of $\alpha$ on $q$ must connect these two regimes.

%W6
\citet{pessah08} examined the dependence of $\alpha$ on $q$
numerically.  They used nonrelativistic shearing box simulations with
resolution $32\times 192 \times 32$ in $r\times \phi\times z$, zero
net magnetic flux, and initial gas-to-magnetic pressure ratio
$\beta=200$.  They ran a series of simulations in which they varied
$q$ from $q=-1.9$ up to $q=1.9$ in steps of $\Delta q = 0.1$.  Each
simulation ran 150 orbits and they measured $\alpha$ by averaging the
data from the last 100 orbits.  For $q>0$, they found the
power-law scaling
\begin{equation}\label{eq:qn}
\alpha \propto q^n,
\end{equation}
where $n$ is between 2 and 8.  Higher
resolution simulations are needed to determine $n$ more precisely.
For $q<0$, they found $\alpha=0$, as expected from the MRI
stability criterion.

%W7
Between the inner edge of a thin accretion disk and the outer,
non-relativistic regions, $q$ only varies by $50\%$ 
(c.f. \S\ref{sec:shear}), but $\alpha$ can vary by much more if
the exponent in equation \eqref{eq:qn} is large.  For $n=8$, the
change in $\alpha$ is a factor of 10.

%W8
To get a quantitative prescription for $\alpha(r)$, we need $q(r)$.
So first we solve standard $\alpha$-disk equations with constant
$\alpha=\alpha_0$.  This gives $q(r)$.  Then we define
$\alpha(r)=\alpha_0 \left[q(r)/1.5\right]^n$.  The exponent is a free
parameter to be determined from MHD simulations.  Now one could
iterate: feed $\alpha(r)$ back into the $\alpha$-disk equations, and
re-evaluate $q(r)$.  For simplicity, we do not iterate.  We check our
model for $\alpha(r)$ against GRMHD simulation data in
\S\S\ref{sec:thesims}-\ref{sec:moresims}, but the simulation data are
too noisy to justify computing $\alpha(r)$ more precisely for now.

The $\alpha$-disk solutions we use are the relativistic
slim disk solutions of \citet{abramowicz1988,sadowski2011}. This is a family of
one-dimensional solutions for black hole accretion with four
free parameters: black hole mass and spin, accretion rate, and
$\alpha$.  At low accretion rates they reduce to the standard thin
disks of \citet{SS73} and \citet{nt73}, and at high accretion rates they become
advection dominated and are similar to slim disks \citep{abramowicz1988,sadowski2011}

This prescription for $\alpha(r)$ neglects the contribution of large
scale, mean magnetic fields, which can exist even in laminar flow.
These are important near and inside the inner edge of the disk, where
the flow acquires a large radial velocity and stretches the frozen-in
magnetic field.  We discuss this contribution to $\alpha(r)$ in the
next section.

\subsection{Mean Magnetic Field Stresses}
\label{sec:meanalpha}

\citet{penna2010} observed large scale, mean field stresses in GRMHD
simulations of black hole accretion disks, and they showed that a one
dimensional model developed by \citet{gammie1999} could fit the data.

%R2
\citet{gammie1999} solved for the motion of a fluid with a frozen-in
magnetic field as it plunges into a Kerr black hole along the
equatorial plane.  The inner boundary of the flow is at the event
horizon and the outer boundary of the flow is at $\rgammie$, where the
flow is assumed to have zero radial velocity and Keplerian angular
velocity.  The governing equations are mass, angular
momentum, and energy conservation, and Maxwell's equations.
There is no dissipation and the pressure and internal energy of the
gas are neglected.  The solutions are time-independent, axisymmetric,
and vertically averaged.  They provide the rest mass density,
velocities, and magnetic field of the flow as a function of radius.
The free parameters are $a/M$, $\rgammie$, and the amount of
magnetic flux threading the horizon.  Following
\citet{penna2010}, we parametrize the flux threading the horizon by
\begin{equation}
\Upsilon = \frac{\int |B^r| dA}{\sqrt{M \dot{M}}},
\end{equation}
which is dimensionless.  A ``magnetically arrested disk'' corresponds
to $\phi_{\rm BH}\equiv \sqrt{\pi} \Upsilon \gtrsim 50$
\citep{narayan2003,tchekhovskoy2011,narayan2012}.

Gammie's code for generating solutions numerically using the shooting
method is available on the web.
\footnote{\tt{http://rainman.astro.illinois.edu/codelib/codes/inflow/src/}}
The solutions are expressed in terms of $B^i$, but
$b^\mu$ follows from equations \eqref{eq:Btob}-\eqref{eq:btoB}, and
the stress, $-b_\hr b_\hp$, follows from
equations \eqref{eq:et}-\eqref{eq:ep}.

\subsection{Combined Model}

Combining the turbulent and mean field contributions to
$\alpha(r)$ gives the one-dimensional prescription
\begin{equation}\label{eq:alphavsr}
\alpha(r) = \alpha_0 \left[\frac{q(r)}{3/2}\right]^n 
          - \alpha_1 \frac{b_\hr(r) b_\hp(r)}{\rho(r)^\Gamma}, \quad (q>0).
\end{equation}
If there are no large scale fields and $q=3/2$, then
$\alpha(r)=\alpha_0$, a constant. If $q<0$, then one should only
include the second term on the RHS.  The \citet{gammie1999} solutions
do not include gas pressure so we have divided the mean field stress
by $\rho(r)^\Gamma$, which is proportional to pressure for a
polytropic gas. This is a crude substitute for the pressure but it
gives an acceptable fit to the simulations discussed below.

Given $M$, $a/M$, $\dot{M}$, $\alpha_0$, $\Gamma$, $\Upsilon$, and
$r_B$, the slim disk equations provide $q(r)$, and the
\citet{gammie1999} equations provide $b_\hr(r), b_\hp(r)$, and
$\rho(r)$.

The remaining free parameters are $\alpha_1$ and $n$.  These
parameters can be inferred from
MHD simulations.  Note that $\alpha_0$ and $n$ control the size and
shape of the turbulent contribution to $\alpha(r)$, and $\alpha_1$ and
$r_B$ control the size and shape of the mean field contribution to
$\alpha(r)$.

\begin{figure}
\includegraphics[width=\columnwidth]{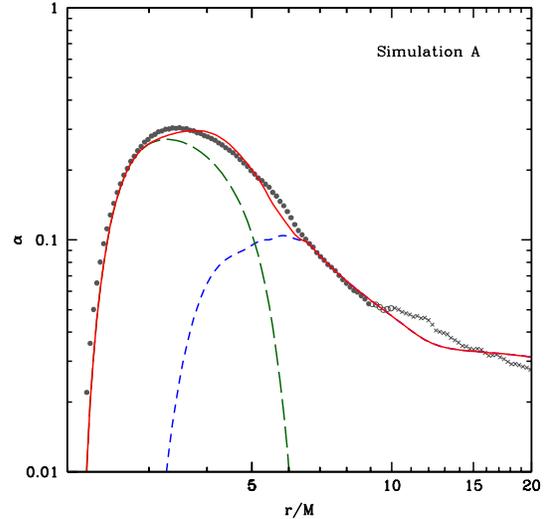}
\caption{The $\alpha(r)$ prescription defined by equation
\eqref{eq:alphavsr}, for parameters $\alpha_0=0.025$, $\alpha_1=100$,
$n=6$, and $\rgammie=6M$ (solid red curve).  This prescription is the
sum of two terms, a mean magnetic field component (long-dashed green),
which dominates inside the ISCO, and a turbulent component (dashed
blue), which dominates outside the ISCO.  These two components are
based on the one-dimensional models of \citet{gammie1999} and
\citet{sadowski2011}, respectively.  At large radii, $\alpha(r)$
converges to $\alpha_0=0.025$, a constant.  This model is a good
description of the data from simulation \thin~(gray points;
c.f. \S\ref{sec:fidsims}).  Data from inside $\rsevere$ are marked
with filled circles, data from between $\rsevere$ and $\rloose$ are
marked with open circles, and data from outside $\rloose$ are marked
with crosses (c.f. \S\ref{sec:simalpha}).}
\label{fig:afitthin}
\end{figure}

%F4
In \S\S\ref{sec:thesims}-\ref{sec:moresims}, we estimate values for
these four parameters by fitting the $\alpha(r)$ prescription to data
from six GRMHD simulations.  As an example, Figure \ref{fig:afitthin}
shows $\alpha(r)$, as prescribed by equation \eqref{eq:alphavsr}, for
the parameters of Model \thin~in Table \ref{tab:afits} (solid red
curve).  The mean magnetic field contribution to $\alpha(r)$
(long-dashed green curve) dominates inside the ISCO, where the
plunging fluid stretches and amplifies the magnetic field.  The
turbulent contribution to $\alpha(r)$ (dashed blue curve) dominates
outside the ISCO, where mean magnetic fields are weak.  At large
radii, the shear parameter becomes $3/2$ and $\alpha$ becomes
constant.  Data from a GRMHD simulation (gray points;
c.f. \S\ref{sec:simalpha}) are in good agreement with the
one-dimensional prescription for $\alpha(r)$.

In the next three sections, we detail
our GRMHD simulations and their connection to the $\alpha(r)$
prescription.

\section{Details of the simulations}
\label{sec:thesims}

\subsection{Computational Method}

%M8
The simulations were carried out with the 3D GRMHD code HARM
\citep{gammie2003,mckinney2006,mckinney2009}, which solves the ideal
MHD equations for the motion of a magnetized gas in the Kerr metric,
the spacetime of a rotating black hole.  The equation of motion of the
gas is taken to be $u=p/(\Gamma-1)$, where $u$ and $p$ are the
internal energy and pressure and $\Gamma$ is the adiabatic index.  The
code conserves energy to machine precision, so any energy lost at the
grid scale by, e.g., turbulent dissipation or numerical reconnection,
is returned to the gas, increasing its entropy.   

%M9
Table \ref{tab:sims} gives a summary of the six simulations, which we
have labeled \thin--\mad.  Simulations \thin, \thinA, and \thinLR~are
thin, radiatively efficient disks, and simulations \sane, \saneA, and
\mad~are thick, radiatively inefficient disks.  The spin parameter is
$a/M=0.7$ for simulations \thinA~and \saneA, and $a/M=0$ for the others.

\begin{table*}
 \centering
 \begin{minipage}{70mm}
  \caption{GRMHD simulation parameters}
  \begin{tabular}{@{}ccccccc@{}}
  \hline
  Simulation
  & $a/M$ 
  & $h/r$
  & Initial loops 
  & \res 
  & $\phi_{\rm max}$
  & Duration\\
  \hline
% 
% \thin   cd /n/home08/rpenna/blackhole/thindisks/hr1a0_V/alpha/
% \thinA  cd /n/home08/rpenna/blackhole/thindisks/hr05a7_IV/alpha/
% \thinLR cd /n/home08/rpenna/blackhole/thindisks/hr05a0_IV/alpha/
% \sane   cd /n/home08/rpenna/blackhole/limotorus_calc/pol-trans-bc/a0-hr03/alpha/
% \saneA  cd /n/home08/rpenna/blackhole/limotorus_calc/pol-trans-bc/a0-hr03/alpha/
% \mad    cd /n/home08/rpenna/blackhole/limotorus_calc/mad/a0/alpha/
%
 \thin   & 0   &  0.1  & Multiple & $256\times128\times64$ & $\pi$ & $20,000M$ \\
 \thinA  & 0.7 &  0.05 & Multiple & $256\times64\times32$  & $\pi/2$ & $27,000M$ \\
 \thinLR & 0   &  0.05 & Multiple & $256\times64\times32$  & $\pi/2$ & $27,000M$ \\
 \sane   & 0   &  0.3  & Multiple & $256\times128\times64$ & $2\pi$  & $200,000M$ \\
 \saneA  & 0.7 &  0.3  & Multiple & $256\times128\times64$ & $2\pi$  & $100,000M$ \\
 \mad    & 0   &  0.3  & Single   & $264\times126\times60$ & $2\pi$  & $100,000M$ \\
\hline \label{tab:sims}
\end{tabular}
\end{minipage}
\end{table*}

Most of the simulations have been described in previous papers, so our
overview of the simulations in this section can be brief.  Simulations
\thinA~and \thinLR~are two of the models discussed by
\citet{kulkarni2011}.  Simulations \thin, \thinA, and \thinLR~were
analyzed by \citet{zhu2012} (where they are labeled E, B, and C,
respectively).  Finally, simulations \sane~and \mad~were studied by
\citet{narayan2012} (where they are called SANE and MAD).  The only
simulation that has not appeared before is \saneA, but it only differs
from \sane~in that it has $a/M=0.7$ and a duration of $100,000M$.

The resolution of simulations \thinA~and \thinLR~in $r\times
\theta\times \phi$ is $256 \times 64\times 32$, and the resolution of
the other simulations is $256\times 128\times 64$.  The radial grid is
logarithmically spaced to concentrate attention on the inner regions
of the flow.  The inner boundary of the grid is between the Cauchy
horizon and event horizon, and outflow boundary conditions are used
there, so the event horizon behaves as a true horizon.  The polar grid
is squeezed towards the equatorial plane to concentrate resolution on
the turbulent, high density regions of the flow, at the expense of the
laminar, coronal regions.  Simulations \sane, \saneA, and \mad~use a
version of the grid developed by \citet{tchekhovskoy2011}, in which
the $\theta$ resolution near the pole increases with increasing radius
so as to follow the formation of jets, which collimate at large
distances.  The azimuthal grid is uniform and extends from 0 to
$\phi_{\rm max}$, where $\phi_{\rm max}$ is either $\pi/2$
(simulations \thin, \thinA, and \thinLR) or $2\pi$ (simulations \sane,
\saneA, and \mad).  

The properties of the six simulations are summarized in Table \ref{tab:sims}.

\subsection{Initial Conditions}

%M11
Initially, the gas orbits the black hole in a torus in hydrostatic
equilibrium \citep{dvhk2003,penna2010,penna2012b}. The thickness of
the torus can be adjusted to give either thin or thick accretion
disks.  A weak poloidal magnetic field threads the torus.  All of the
simulations start with a sequence of poloidal loops, except \mad,
which starts from a single magnetic loop.  When there are multiple
loops, the black hole accretes flux of alternating polarity over time
and little net flux builds up on the hole.  In simulation \mad, the
center of the loop at $r=300M$ does not reach the black hole over the
duration of the simulation, so the black hole acquires a large net
flux.  In all of the simulations, the magnetic field is normalized so
that the initial gas-to-magnetic pressure ratio has minimum
$\beta=100$.

\subsection{Quasi-Steady State}

The initial condition is unstable to the MRI.  Differential rotation
of the torus triggers the MRI and the gas becomes turbulent after
$\sim 10$ orbits.  Turbulence transports angular momentum and energy
outwards and the gas accretes inwards.  At late times in the
simulation, gas in the disk near the midplane is turbulent.  Magnetic
buoyancy lifts fields above and below the disk,
forming a highly magnetized corona.  The corona is mostly laminar
because the MRI requires $\beta >1$.

\begin{figure*}
\includegraphics[width=0.49\linewidth]{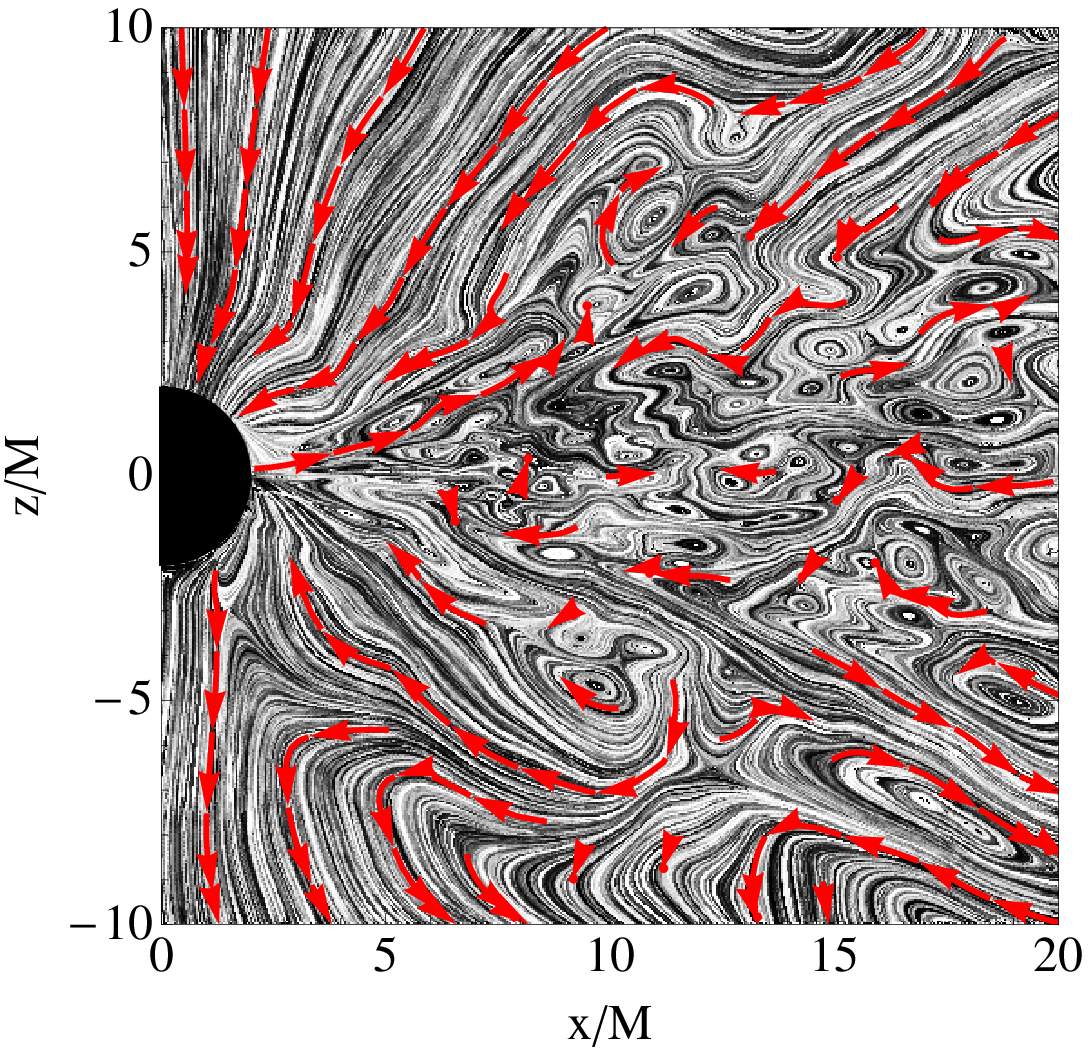}
\includegraphics[width=0.49\linewidth]{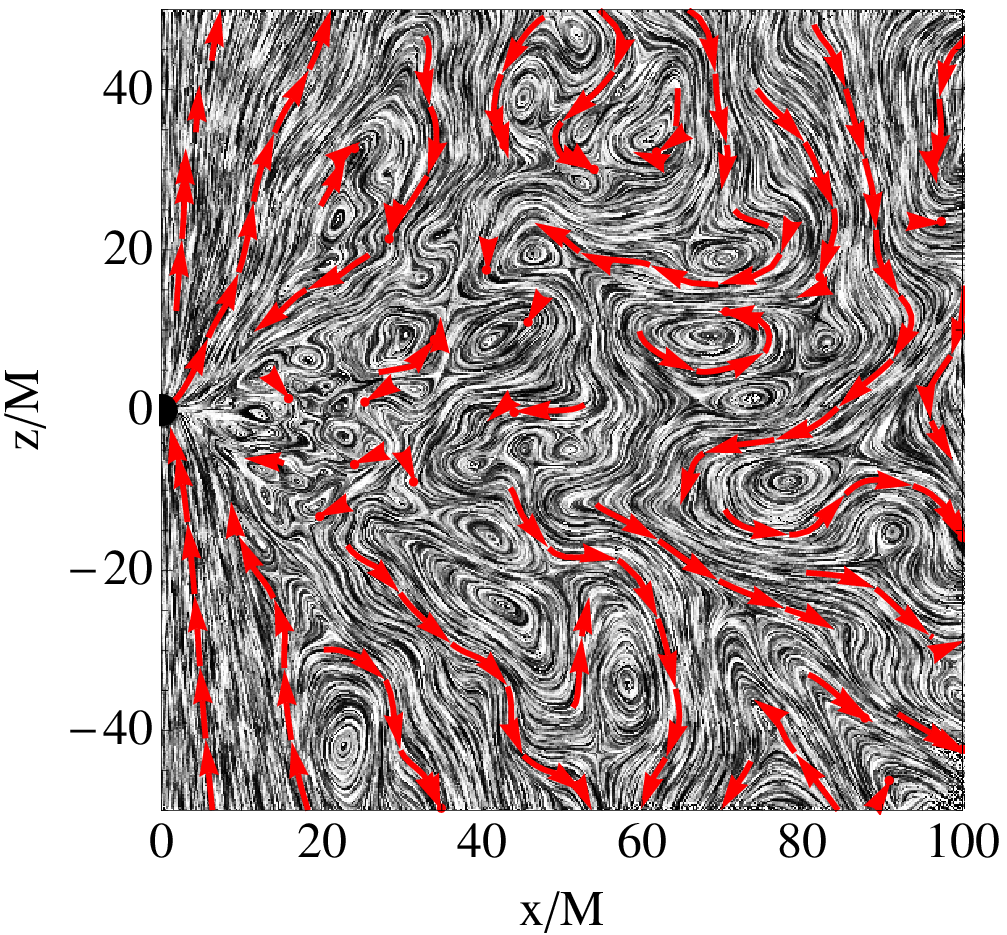}
\caption{\emph{Left panel:} Black, white, and red streamlines show the
  poloidal, fluid frame magnetic field $(b^\hr,b^\hh)$ for simulation
  \thin.  The field has been azimuthally averaged to make the
  streamlines appear continuous in this two-dimensional projection
  (this makes the flow appear slightly less turbulent).
  Turbulent twisting of the magnetic field can be seen on different
  scales.  When the field is twisted on the smallest scale, the
  grid scale, there is reconnection and dissipation.  The density
  scale height of the disk is $h/r \sim 0.1$ (c.f. 
  \S\ref{sec:hr}).  Fluid in the disk region of the
  flow is turbulent.  Magnetic buoyancy lifts magnetic fields out of
  the disk where they settle in a highly magnetized coronal region.
  The coronal region is mostly laminar, because magnetic tension is
  quenching the MRI.  On very large scales, the magnetic field has an
  approximately dipolar structure. \emph{Right panel:} Same as left
  panel, but for simulation \sane.  This accretion flow is much thicker
  and the flow only becomes laminar near the polar axes.}
\label{fig:stream}
\end{figure*}

Figure \ref{fig:stream} shows the fluid frame magnetic field at the
end of simulations \thin~and \sane.  The field has been azimuthally
averaged but not time averaged.  The coordinates are $x/M=r
\sin(\theta)$ and $z/M=r\cos(\theta)$.  The turbulent region of
simulation \thin~is thinner than the turbulent region of simulation
\sane.  The turbulent region of simulation \sane~extends nearly to the
polar axes.

\section{Analysis of Simulations \thin~and \sane}
\label{sec:fidsims}

In this section we discuss our analysis of simulations \thin~and
\sane. These describe a prototypical thin disk and
a prototypical thick disk around non-spinning black holes.  We discuss
the remaining four simulations in \S\ref{sec:moresims}

Our goal for this section, achieved in \S\ref{sec:simfits}, is to
extract $\alpha(r)$ profiles
from the simulation data and compare them with the one-dimensional
prescription defined by equation \eqref{eq:alphavsr}.  As a first
step, we discuss the distinction between disk and coronal fluid.  We
only include disk fluid in our calculations.  Then we discuss the
radial range of fluid that can be considered to have reached a
quasi-steady state.  We only include quasi-steady data in our
calculations.  We examine the shear rate and epicyclic frequency of the
simulations, because these play an important role in determining
$\alpha$.  Finally, we compute $\alpha(r)$ and compare it with our
prescription from \S\ref{sec:1dalpha}

\subsection{The Distinction Between Disk and Corona}
\label{sec:hr}

We would like to separate the disk component of the flow from the
coronal component so that we can focus our analysis on the disk.
There are several reasons to do this.  For one, the stress has a
different character in the corona and disk regions of the flow.  In
the corona, the stress is mostly generated by mean magnetic fields,
while in the disk, the stress is generated largely by turbulence.  So
including coronal stresses in the model would add new difficulties.
For this reason, and also for simplicity, we focus on the disk region
of the flow.

There are other reasons to isolate the disk from the corona.  At least in
thin accretion disks, the emission from the corona and disk are
different.  The disk has a thermal spectrum and the corona has a power
law spectrum.  So the distinction is sensible for observations.
Accretion disk models which use the $\alpha$-viscosity prescription
tend to focus on the disk region of the flow and ignore the corona.
Another reason to separate out the disk region is that our numerical
grid concentrates $\theta$ resolution at the midplane and leaves
the polar regions poorly resolved.  So the simulation data are
unreliable in the corona.

%Sat1
We therefore only include fluid within one density scale height
of the midplane in our analysis.  The density scale height is defined as,
\begin{equation}
\frac{h}{r} = \frac{
     \int_0^{2\pi}  \int_0^{\phi_{\rm max}} \int_{t_1}^{t_2}
        \left| \theta-\pi/2 \right| 
          \rho u^t \sqrt{-g} dt d\theta d\phi}
     {\int_0^{2\pi}  \int_0^{\phi_{\rm max}} \int_{t_1}^{t_2}
          \rho u^t \sqrt{-g} dt d\theta d\phi}.
\end{equation}
where $\rho$ is rest mass density in the fluid frame, and $\rho u^t$ is
rest mass density in Boyer-Lindquist coordinates.  The time integral
is over the steady state period of the flow, as explained below.
Another popular definition for the scale height is
$\left(h/r\right)_{\rm rms}=\left(\int \left(\theta-\pi/2\right)^2
\rho \sqrt{-g} dtd\theta d\phi /\int \rho \sqrt{-g} dt d\theta
d\phi\right)^{1/2}$.  We have no reason to favor one definition over
another, though it should be noted that $(h/r)_{\rm rms}$ can be
a factor of $\sim 2$ bigger than $h/r$ \citep{penna2010}.

We also add a density-weighting to vertical averages, to further
emphasize midplane fluid.  That is, the density weighted vertical
average of $\mathcal{O}$ is $\int \mathcal{O} \rho u^t \sqrt{g_{\theta\theta}}d\theta /
\int \rho u^t \sqrt{g_{\theta\theta}}d\theta$.

\begin{figure}
\includegraphics[width=\columnwidth]{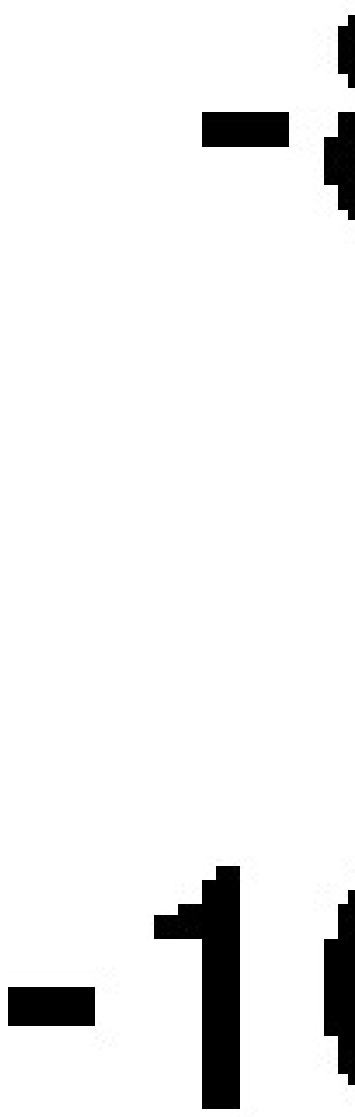}\\
\includegraphics[width=\columnwidth]{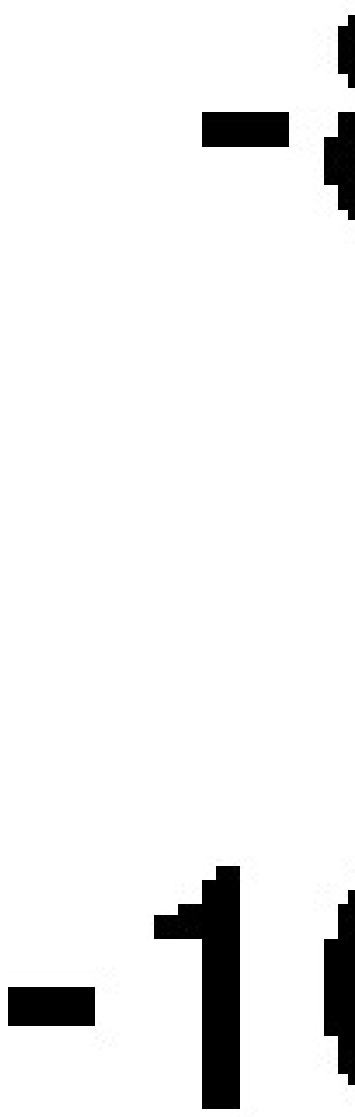}
\caption{\emph{Top panel:} $\log(\alpha)$ in the $r-\theta$ plane for
  simulation \thin.  The data has been time-averaged over
  $t=7,000M-20,000M$.  Dashed black lines indicate one density scale
  height above and below the midplane (c.f. \S\ref{sec:hr}).  We refer
  to the low $\alpha$ region within one scale height of the midplane
  as the disk, and the high $\alpha$ region outside one scale height
  as the corona. We restrict our calculations to the disk, for the
  reasons discussed in \S\ref{sec:hr}.  We show this plot here to
  illustrate the difference between the disk and the corona.  We
  explain how $\alpha$ is obtained from the simulation in
  \S\ref{sec:simalpha}. \emph{Bottom panel:} Time-averaged ratio of
  Maxwell stress to Reynolds stress on a log scale, in the $r-\theta$
  plane, for simulation \thin.  Coronal fluid is more magnetically
  dominated than disk fluid.  This is expected, as magnetic buoyancy
  lifts magnetic fields into the corona.  }
\label{fig:alpha2dthin}
\end{figure}

%Sat2
The top panel of Figure \ref{fig:alpha2dthin} shows $\log(\alpha)$ as
a function of $x$ and $z$ for simulation \thin.  We explain how
$\alpha$ is obtained from the simulation in \S\ref{sec:simalpha}, but
we show this plot here to illustrate the distinction between the disk
and the corona. Black dashed curves mark one scale height above the
midplane.  Note that $\alpha$ has a very different character above and
below the disk region.  In the coronal regions $\alpha$ is much larger
than in the disk.  The bottom panel of Figure \ref{fig:alpha2dthin}
shows the ratio of Maxwell stress to Reynolds stress on a logarithmic
scale.  Maxwell stresses are much more significant in the corona.  The
different character of the flow in these two regions is one of the
reasons we only include disk fluid within one $h/r$ of the midplane in
our calculations.

\begin{figure}
\includegraphics[width=\columnwidth]{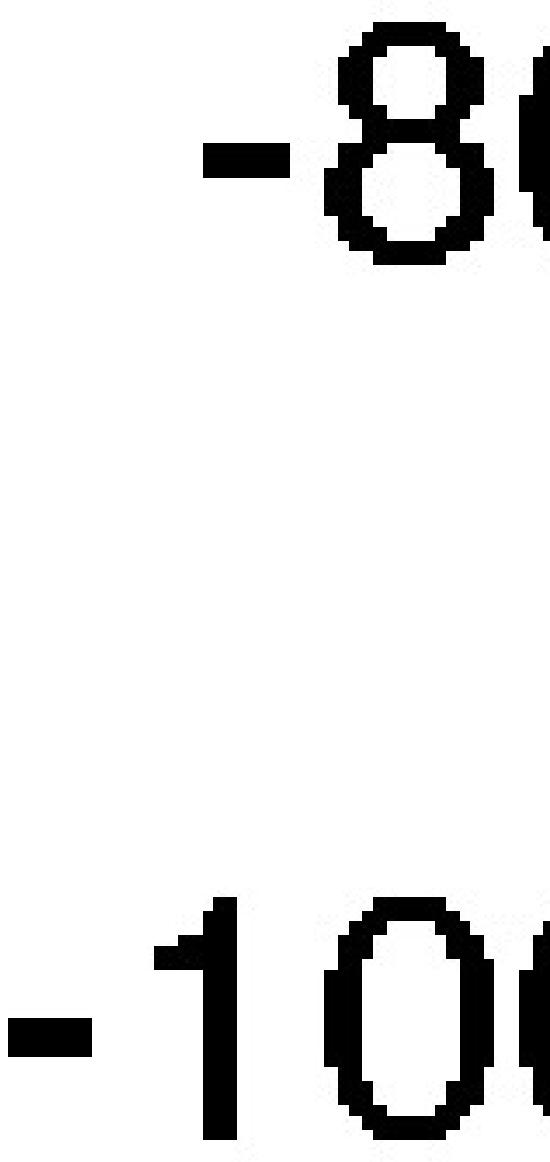}\\
\includegraphics[width=\columnwidth]{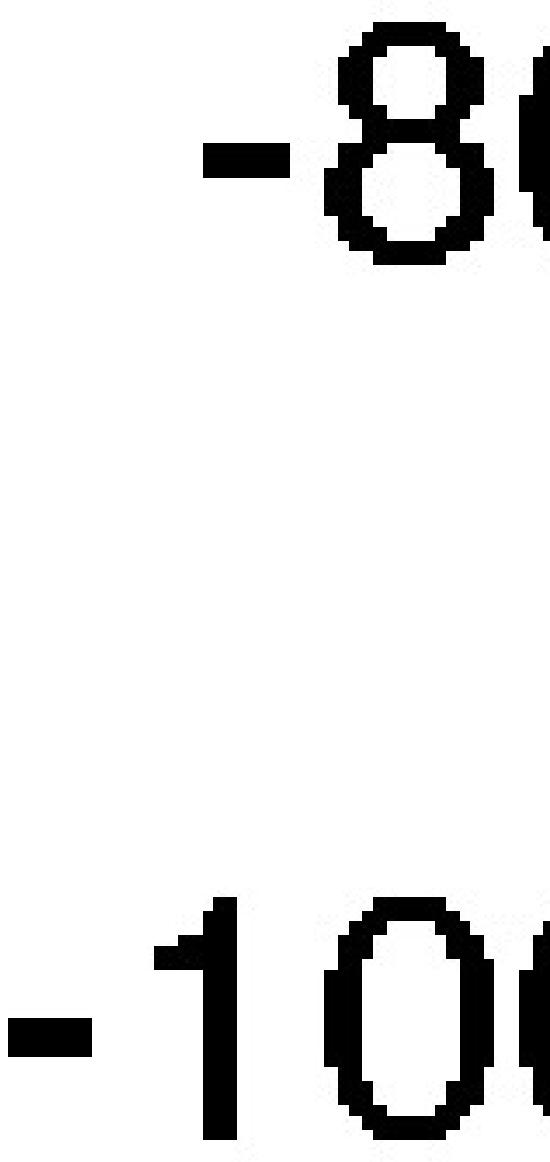}
\caption{Same as Figure \ref{fig:alpha2dthin}, except for simulation
  \sane.  Unlike simulation \thin, the highly magnetized region does
  not have the same shape as the disk scale height.  In fact, the
  former has a paraboloidal, jet-like shape, approximately
  $z/M=\left(r/6M\right)^2$.}
\label{fig:alpha2dsane}
\end{figure}

%Sat3
Figure \ref{fig:alpha2dsane} shows the same quantities for simulation
\sane.  Again, $\alpha$ and the Maxwell stress are much larger outside
the disk.  However, the shape of this region does not track $h/r$ as
it did in simulation \thin.  In fact, the high $\alpha$, high Maxwell
stress region has a parabolic shape, bounded by roughly
$z/M=\left(r/6M\right)^2$.  This looks like a jet.  For simplicity and consistency, we
will restrict our calculations to the fluid within one $h/r$ of the
midplane for all of the simulations.  We have checked that our results
do not depend on the details of this cutoff, as long as 
we do not include the ``jet'' region.

\subsection{Radial and Azimuthal Velocities}

%Sat4
To extract smooth results from turbulent data, it is necessary to
average the data over at least a viscous time, which smooths out
turbulent fluctuations \citep{narayan2012}.  As a first step, we need
the radial velocity of the simulations.  To compute the shear rate and
epicyclic frequency of the gas (and thus $\alpha$), we need the
azimuthal velocity of the gas.  We compute these two components of the
velocity in this section.  We will not need the $\theta$ component of
the velocity.  It is much smaller than the radial and azimuthal
components in the disk region of the flows.

%Sat5
In a thin accretion disk, the gas closely follows circular geodesics
as it spirals toward the ISCO.  The radial velocity of the gas is
significantly smaller than the azimuthal velocity.  In a standard thin
disk, the radial velocity is suppressed by a factor of $\alpha
(h/r)^2$.  In the ADAF solution, which describes a very thick
accretion flow, the radial velocity is suppressed by a factor of
$\alpha$ relative to the azimuthal velocity.  These relations hold
approximately for the GRMHD simulations as well.

%Sat6
%In Boyer-Lindquist coordinates, the radial velocity of the gas is zero
%at the horizon.  This is because Boyer-Linquist coordinates are ``at
%rest with respect to infinity,'' and gravitational redshift causes
%motion at the horizon to appear to freeze.  The velocity that is
%relevant to the physics of turbulence is a locally measured radial
%velocity.  Of course, the radial velocity in the fluid frame itself is
%zero.  
The radial velocity most relevant for turbulence is the one measured
by the zero angular momentum observer (ZAMO) of \citet{bardeen1972}.
This is a local, inertial frame attached to observers with zero
angular momentum.  As a result of frame dragging, ZAMO observers
appear to rotate with respect to observers at infinity.  The ZAMO
frame is at rest with respect to the local spacetime.

%Sat7
The radial velocity in the ZAMO frame is
\begin{equation}
v_r = \frac{\sqrt{A}}{\Delta}\frac{u^r}{u^t},
\end{equation}
where $\Delta$ and $A$ are defined by equations \eqref{eq:delta} and \eqref{eq:A}.

\begin{figure*}
\includegraphics[width=0.49\linewidth]{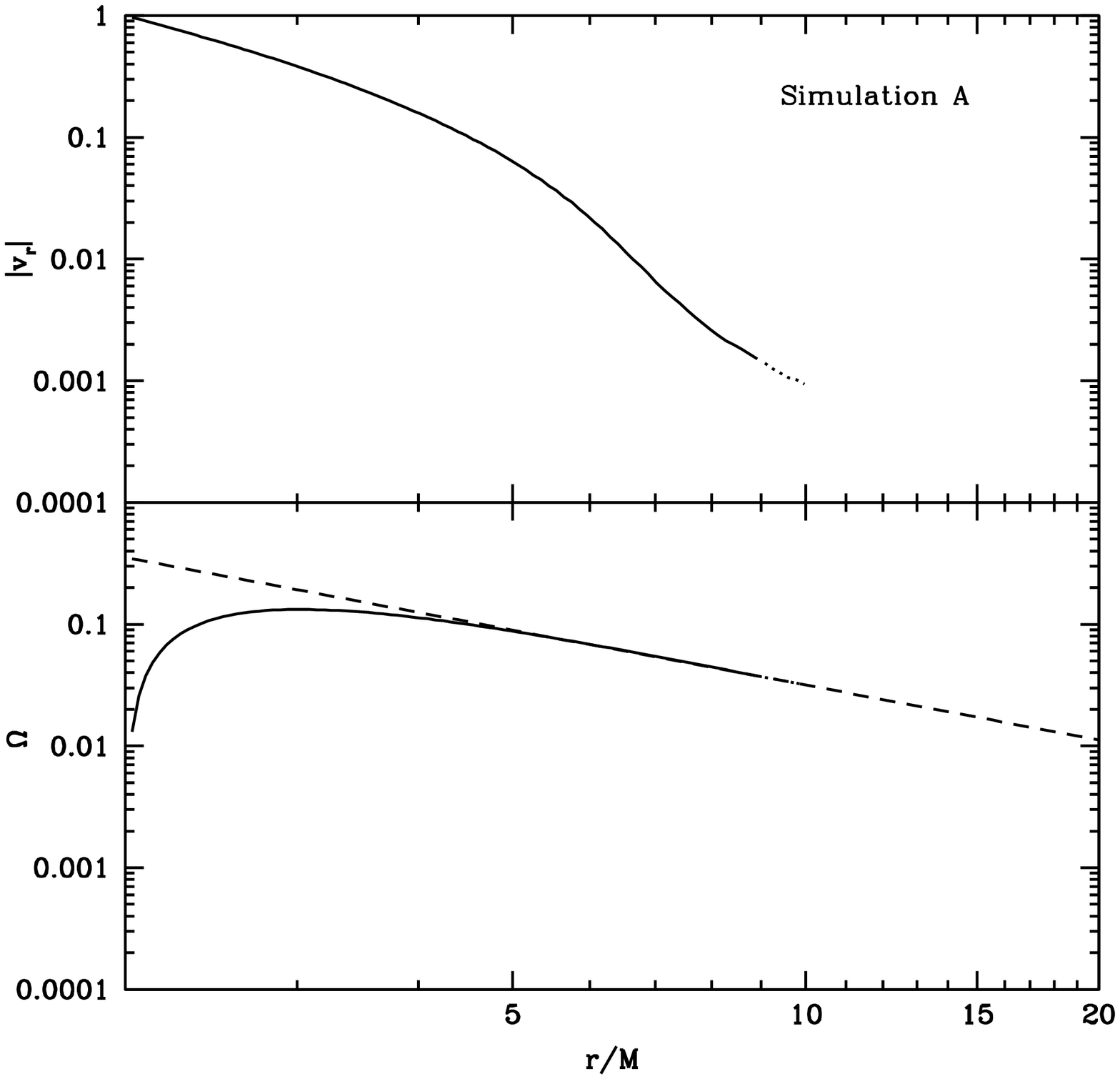}
\includegraphics[width=0.49\linewidth]{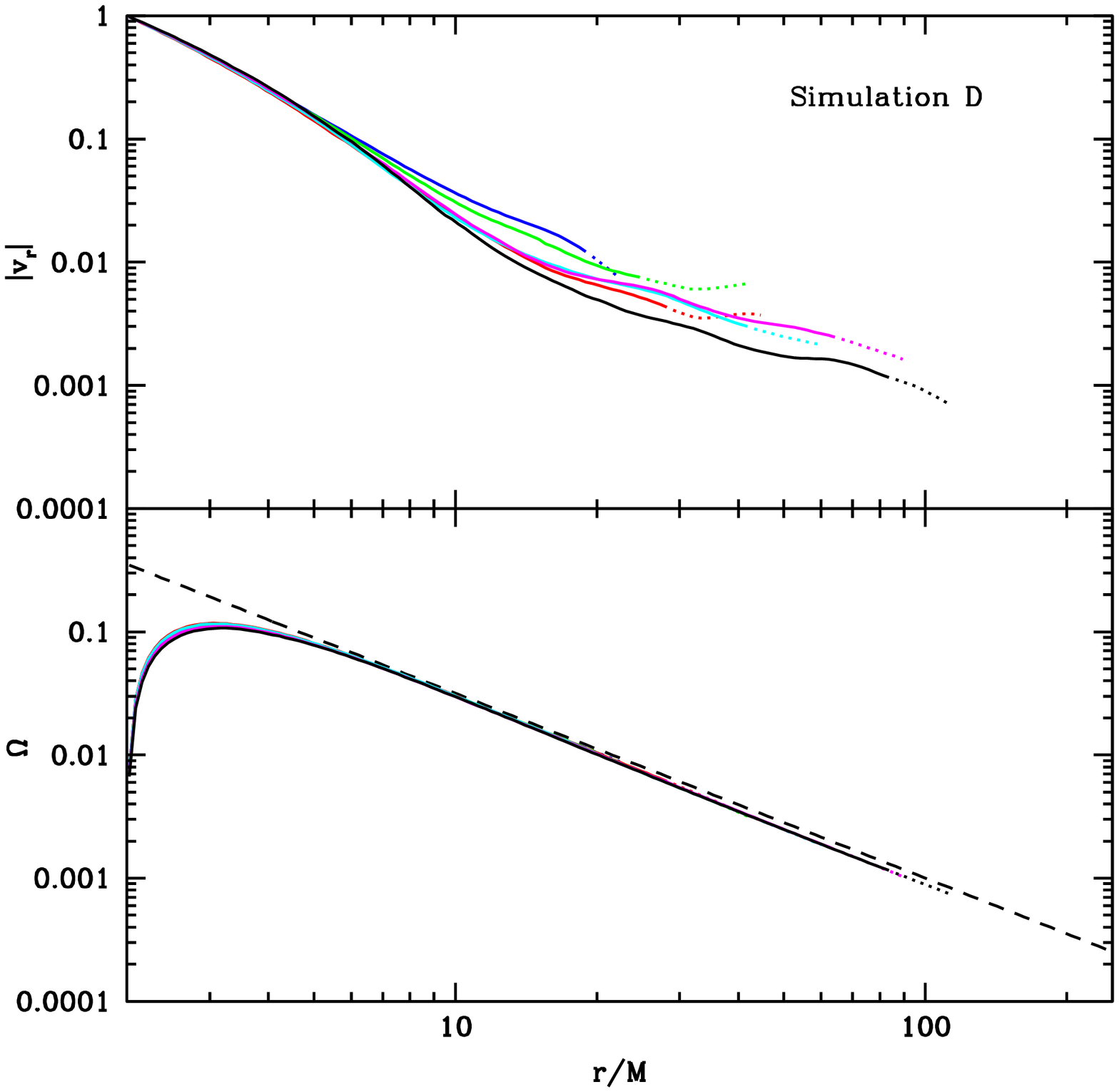}
\caption{
\emph{Top left:} Radial velocity as a function of radius for
simulation \thin.  The solid line extends to $r=\rsevere$ and the
dashed line extends to $r=\rloose$ (estimated convergence radii of
\S\ref{sec:steady}).  The radial velocity increases suddenly around
the ISCO at $r=6M$, inside of which there are no stable circular
orbits for the gas to follow.
\emph{Top right:} Radial
velocity as a function of radius for simulation \sane.  Colors
correspond to time chunks 1 (blue), 2 (green), 3 (red), 4 (cyan), 5
(magenta), and 6 (black) (see \S\ref{sec:steady}).
\emph{Bottom left:} Angular velocity as a function of radius for
simulation \thin.  The dashed black curve shows the angular velocity
of circular, equatorial geodesics.  The simulated flow is nearly geodesic
outside the ISCO.  The angular velocity has a maximum near the photon
orbit at $r=3M$, so the shear parameter (and hence the turbulent
contribution to $\alpha$) will be zero here.
\emph{Bottom right:} Angular velocity as a function of radius for
simulation \sane.  This flow is slightly sub-Keplerian outside the
ISCO.
}\label{fig:vrthin}
\end{figure*}

The top left panel of Figure
\ref{fig:vrthin} shows the radial velocity as a function of radius for
simulation \thin.  The data has been averaged over the disk region of
the flow, with a density weighting, as discussed in \S\ref{sec:hr}.
It has been time-averaged over the quasi-steady state period of the
flow, which we explain below.  The ISCO is at $r=6M$.  Inside the
ISCO, the gas is approximately in free fall and the radial velocity
increases rapidly as the gas approaches the black hole.  Outside the
ISCO, the motion is more nearly circular.  In this region, the radial
velocity is suppressed relative to the azimuthal velocity as predicted
by standard disk theory.  Because the radial velocity is small, it
takes a long time for the simulation to reach a quasi-steady state.
Once steady state is reached, it requires a time average extending
over many orbital periods to smooth out turbulent fluctuations and
obtain reliable results.  We take up these issues in the next section.

%Sat8
The top right panel of Figure \ref{fig:vrthin} shows the radial
velocity as a function of radius for simulation \sane.  The radial
velocity is much larger than the radial velocity of \thin, because
this accretion flow is geometrically thick.  For this reason, the data
from this simulation is in quasi-steady state out to a larger
radius.  Also, the larger radial velocity smears out the inner
edge of the accretion disk.  There is no longer a sudden transition
between slow and fast radial velocity, at the ISCO or at any other
radius.

%Sat9
The bottom left panel of Figure \ref{fig:vrthin} shows the angular
velocity
\begin{equation}
\Omega = \frac{d\phi}{dt}=\frac{u^\phi}{u^t},
\end{equation}
of simulation \thin~as a function of radius.  Again, we have included
only fluid within one scale height of the midplane, taken the
density-weighted vertical average, and time averaged over the steady
state portion of the flow.  The dashed line shows the angular velocity
of circular geodesics in the equatorial plane, $\Omega = M^{1/2} /
(r^{3/2}+a M^{1/2})$.  The flow follows circular geodesics except for
$r \lesssim 5M$, where the radial velocity is increasing rapidly.  The
angular velocity peaks around $3M$.  At this radius, the shear
parameter $q$ must go to zero, so the turbulent contribution to $\alpha$
becomes negligible.

%Sat10
The bottom right panel of Figure \ref{fig:vrthin} shows the angular
velocity of simulation \sane.  It is also nearly Keplerian outside the
ISCO.  This is perhaps surprising, because the angular velocity of the
self-similar ADAF solution is very sub-Keplerian.  The initial torus
of the GRMHD simulation persists at large radii over the duration of
the simulation and continues to feed nearly Keplerian gas into the
inner regions of the flow.  This acts as a very strong boundary
condition, which may limit the solution's ability to converge to the
ADAF solution.  The ADAF solution is self-similar and describes an
accretion flow with infinite extent but, as we will see, the simulation
is only converged out to $r=100M$.  A longer duration simulation,
which has reached quasi-steady state out to a larger radius, might be
expected to have a more ADAF-like angular velocity.  Nonetheless, the radial
velocity of simulation \sane~does appear to be converging to the ADAF
prediction, as shown by \citet{narayan2012}.

\subsection{Convergence and Steady State}
\label{sec:steady}

Following \citet{narayan2012}, we divide the data from simulation
\sane~into six ``time chunks'' which are logarithmically spaced in
time.  Each time chunk is about twice as long as the previous one.
They are summarized in Table \ref{tab:sane}.  This logarithmic spacing
is useful since most of the quantities we are interested in show
power-law behavior as a function of both time and radius.  Note that
there is no overlap between chunks, and hence each chunk provides
independent information.  Because the duration of simulation \thin~is
only $20,000M$, we use a single time chunk, spanning $t=7,000M-20,000M$.

For each time chunk, we compute the time-averaged radial velocity
profile $v_r(r)$ of the gas within one scale-height of the mid-plane.
We estimate the viscous time at radius $r$ by \citep{nt73,penna2010}:
\begin{equation}
\tacc(r) = \frac{r}{|v_r(r)|}.
\end{equation}
Following \citet{narayan2012}, we then define two criteria, one
``strict'' and one ``loose,'' to estimate the radius range over which
the flow has achieved inflow equilibrium:
\begin{align}
\tacc(\rsevere) &= \tchunk/2, \\
\tacc(\rloose) &= \tchunk,
\end{align}
where $\tchunk$ is the duration of the chunk.  The values of
$\tchunk$, $\rsevere$, and $\rloose$ for the various time chunks are
summarized in Tables \ref{tab:thin} and \ref{tab:sane}.  We only trust
data from inside $\rloose$.  Data from inside $\rsevere$ are
considered particularly reliable.

\begin{table*}
 \centering
 \begin{minipage}{70mm}
  \caption{Convergence radii for simulations \thin, \thinA, and \thinLR}
  \begin{tabular}{@{}ccccc@{}}
  \hline
   Simulation & Time Range (M) & $\tchunk/M$ & $\rsevere/M$ & $\rloose/M$\\
 \hline
 \thin   &  7,000-20,000      & 13,000    & 9 & 10  \\
 \thinA  & 20,000-27,000      &  7,000    & 6.5 & 7 \\
 \thinLR & 20,000-27,000      &  7,000    & 6.5 & 7  \\
\hline \label{tab:thin}
\end{tabular}
\end{minipage}
\end{table*}

\begin{table*}
 \centering
 \begin{minipage}{70mm}
  \caption{Time chunks for simulation \sane}
  \begin{tabular}{@{}ccccc@{}}
  \hline
  Chunk & Time Range (M) & $\tchunk/M$ & $\rsevere/M$ & $\rloose/M$\\
 \hline
 I   & 3,000-6,000      & 3,000    & 19 & 23 \\ %blue
 II  & 6,000-12,000     & 6,000    & 25 & 43 \\ %green
 III & 12,000-25,000    & 13,000   & 29 & 45 \\ %red
 IV  & 25,000-50,000    & 25,000   & 43 & 62 \\ %cyan
 V   & 50,000-100,000   & 50,000   & 66 & 92 \\ %magenta
 VI  & 100,000-200,000  & 100,000  & 86 & 113\\ %black
\hline \label{tab:sane}
\end{tabular}
\end{minipage}
\end{table*}

\subsection{Shear and Epicyclic Frequencies}

%Sun1
Now we compute the dimensionless shear parameter, $q$, and the
epicyclic frequency, $\kappa$, as a function of radius.  The shear
parameter is computed from $q=-2\shear/\Omega$.  The epicyclic frequency is
computed from the shear parameter using equation \eqref{eq:epicyclic}.  The data are
vertically averaged over the gas within one scale height of the
midplane using the density weighting.  The data are time averaged over
the time chunks listed in Tables \ref{tab:thin} and \ref{tab:sane}.

\begin{figure*}
\includegraphics[width=0.49\linewidth]{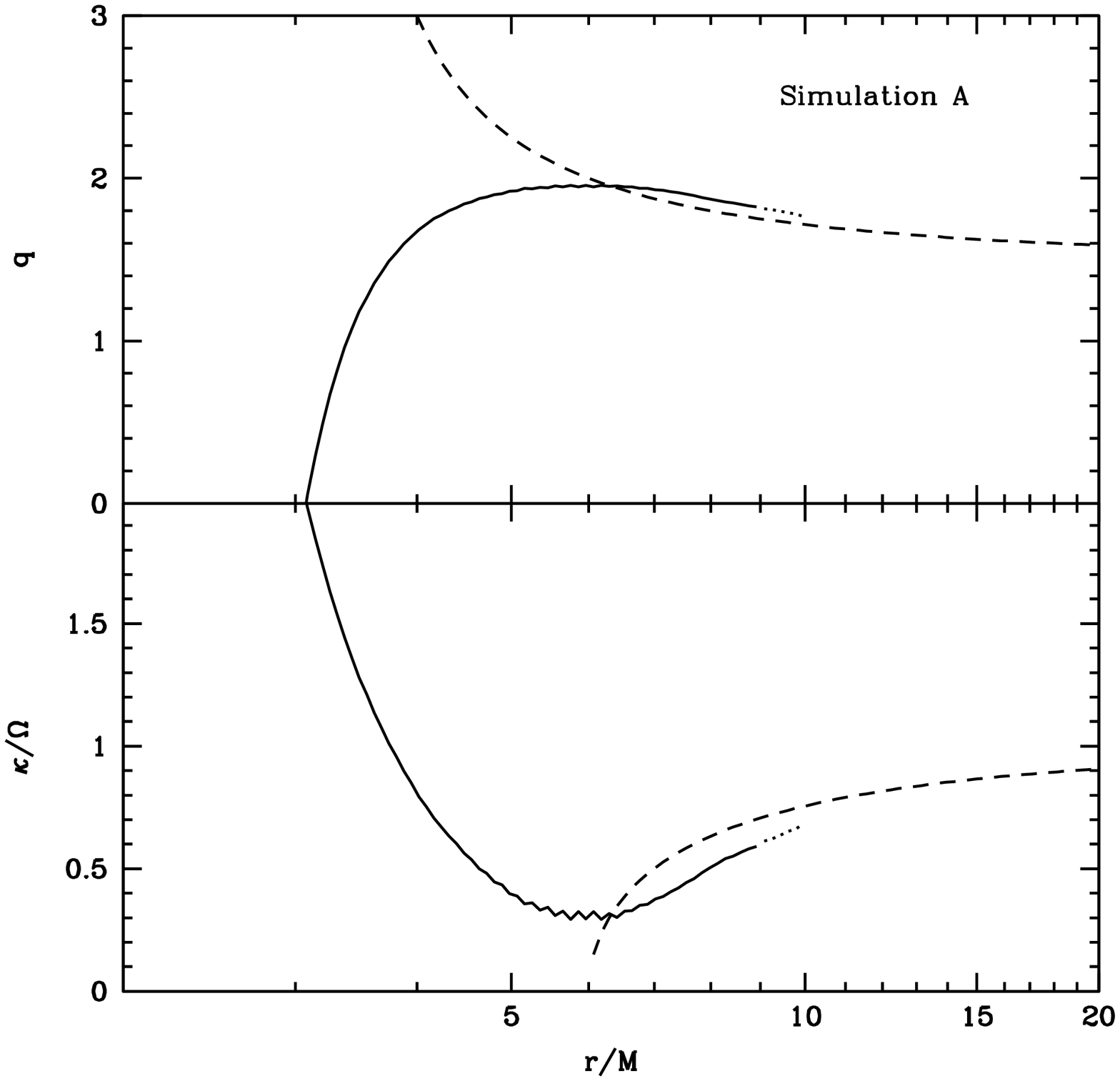}
\includegraphics[width=0.49\linewidth]{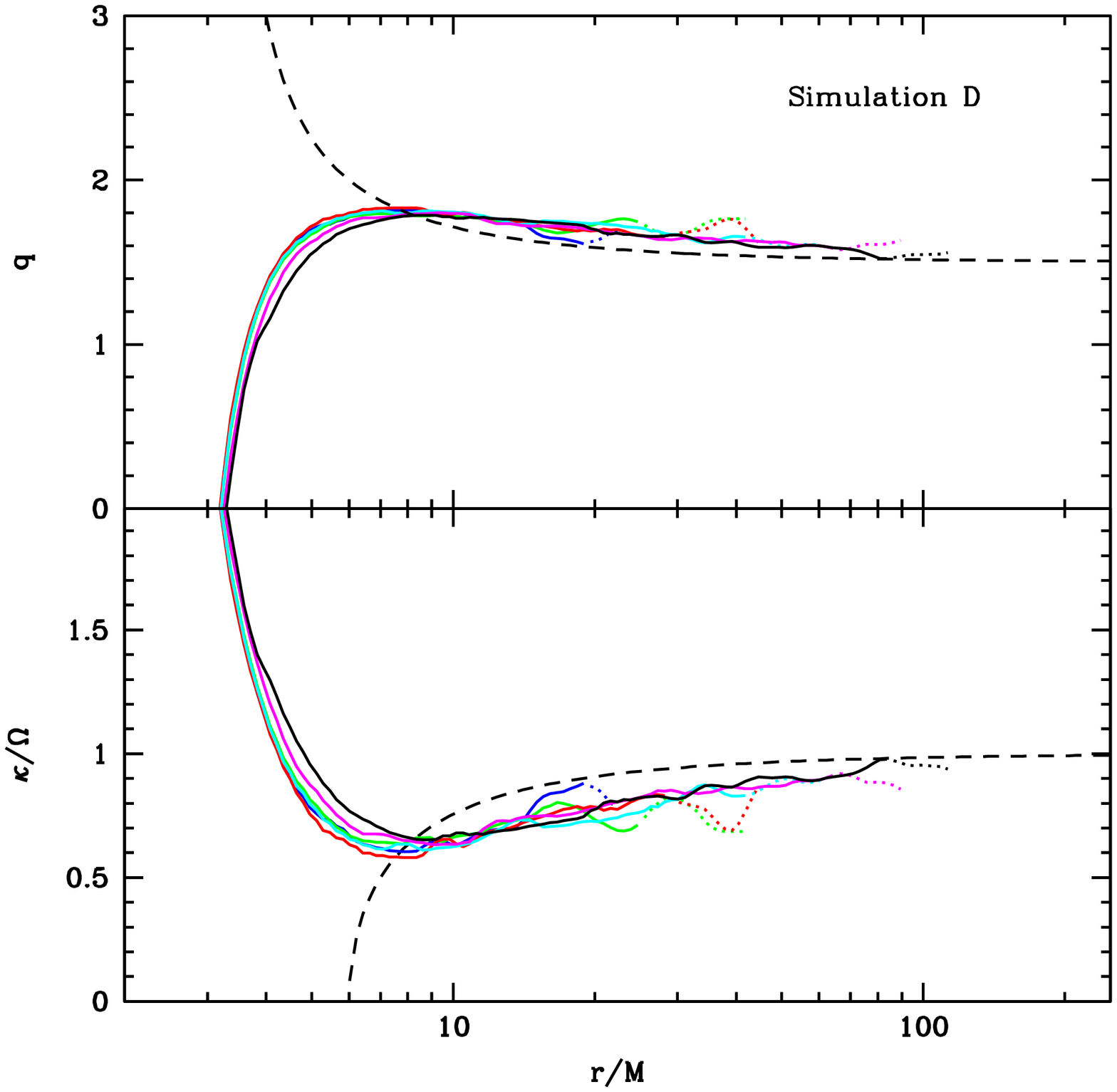}
\caption{
\emph{Top left:} Dimensionless shear parameter, $q$, as a function of
radius for simulation \thin~(solid and dotted curves) and for
Keplerian flow (dashed).  The simulation data are plotted out to
$\rloose$ (dotted curves) and out to $\rsevere$ (solid curves).  The
simulated shear parameter turns over near the ISCO.
\emph{Top right:} Dimensionless shear parameter as a function of
radius for simulation \sane.  Colors are as in Figure \ref{fig:vrthin}.
\emph{Bottom left:} Epicyclic frequency as a function of radius for
simulation \thin.  The epicyclic frequency has its minimum near the
ISCO, as expected.  
\emph{Bottom right:} Epicyclic frequency as a function of radius for
simulation \sane.  The minimum is still near the ISCO, but it is
broader and shallower than the minimum in the epicyclic frequency of
simulation \thin.  The larger radial velocity of simulation \sane~has
``smeared out'' the inner edge of the disk.
}
\label{fig:qthin}
\end{figure*}

%Sun2
The top left panel of Figure \ref{fig:qthin} shows the shear parameter
as a function of radius for simulation \thin~out to $\rsevere$ (solid
curve) and $\rloose$ (dotted curve).  The analytical shear parameter
for circular, equatorial Kerr geodesics is shown for comparison
(dashed curve).  At large radii, the analytical $q$ converges to the
shear parameter of non-relativistic Keplerian flow, $q=3/2$.  At the
ISCO, general relativistic corrections increase the shear parameter to
$q=2$.  The GRMHD shear parameter is about $10\%$ larger than the
analytical shear parameter outside the ISCO.  Inside the ISCO, the
analytical $q$ blows up as it approaches the photon orbit. The GRMHD
shear parameter goes to zero near the photon orbit and is negative
very close to the black hole.

%Sun3
The top right panel of Figure \ref{fig:qthin} shows the shear
parameter as a function of radius for simulation \sane.  Results are
shown for each of the time chunks.  All of the time chunks are
consistent out to $\rloose$ to within several percent.  This gives us
confidence that the simulation has converged to a quasi-steady
solution.  The GRMHD shear parameter is similar to the shear parameter
of simulation \thin.  It is about $10\%$ larger than the analytical
$q$ outside the ISCO, turns over inside the ISCO, and drops to zero
near the photon orbit.  There is good agreement between the GRMHD and
analytical shear parameters out to $\rloose \sim 100M$.

%Sun4
The bottom left panel of Figure \ref{fig:qthin} shows the epicyclic
frequency as a function of radius for simulation \thin.  The bottom
right panel shows the same quantity for simulation \sane.  In both
cases, outside the ISCO, the epicyclic frequency of the simulation is
about $10\%$ lower than the epicyclic frequency of Keplerian flow.  In
both cases the epicyclic frequency has a minimum near the ISCO.  The
minimum of the epicyclic frequency roughly marks the most unstable
radius in the flow, because $\kappa=0$ corresponds to marginal
stability.  So it is consistent with standard disk theory that the
minimum of $\kappa$ is near the ISCO.  Simulation \sane~has a broader
and shallower minimum, indicating the inner edge of this disk has been ``smeared
out'' by the larger radial velocity of the flow.

\subsection{Shakura-Sunyaev Viscosity Parameter, $\alpha$}
\label{sec:simalpha}

Finally, we compute the dimensionless viscosity parameter, $\alpha$.
The GRMHD stress-energy tensor is a combination of Reynolds and
Maxwell terms:
\begin{equation}
T_{\mu\nu}=T^{\rm (rey)}_{\mu\nu}+T^{\rm (mag)}_{\mu\nu},
\end{equation}
where,
\begin{align}
T^{\rm (rey)}_{\mu\nu} &= (\rho+u) u_\mu u_\nu +p h_{\mu\nu},\\
T^{\rm (mag)}_{\mu\nu} &= \frac{1}{2}
  \left( b^2 u_\mu u_\nu 
 +b^2 h_{\mu\nu}-2b_\mu b_\nu\right),
\end{align}
and $h_{\mu\nu}=g_{\mu\nu}+u_\mu u_\nu$ is the projection tensor.  To
each term there is an associated stress, which is the $r\phi$
component of the tensor measured in the fluid frame.  So the Reynolds
stress is $\trey_{\hr\hp}$ and the Maxwell stress is $\tmax_{\hr\hp}$.
An important difference between the two is that the Reynolds stress
requires turbulence whereas the Maxwell stress can be
generated by turbulence or large scale magnetic fields and thus can be
nonzero even in laminar flow.

We define $\alpha$ as the ratio of total stress to total pressure:
\begin{equation}
\alpha = \frac{\trey_{\hr\hp}+\tmax_{\hr\hp}}{p+b^2/2}.
\end{equation}
We have chosen to include $b^2/2$ in the denominator because it keeps
$\alpha<1$.  As a result of this choice, part of the observed
dependence of $\alpha$ on $q$ is inherited via the magnetic pressure
because the magnetic pressure is amplified by shear.  This is
significant inside the ISCO where the magnetic pressure is comparable
to or exceeds the gas pressure.

%Sun6
We compute the Maxwell and Reynolds stresses in the rest frame
of the mean flow,
\begin{equation}\label{eq:meanu}
\bar{u}^\mu = \frac{1}{\phi_{\rm max}} \int_0^{\phi_{\rm max}} u^\mu d\phi,
\end{equation}
rather than in the rest frame of the instantaneous flow, $u^\mu$.  The
mean flow is the $\phi$-averaged instantaneous flow.  There are some
subtleties in the distinction between these two frames.  The Reynolds
stress vanishes in the rest frame of the instantaneous flow, $u^\mu$,
by definition of the fluid frame, but not in the rest frame of the
mean flow, $\bar{u}^\mu$.  This is our primary reason for choosing
$\bar{u}^\mu$ to define the inertial fluid frame when computing
$\alpha$ .  The electric field vanishes in the rest frame of the
instantaneous flow, by the assumption of ideal MHD, and not in the
rest frame of the mean flow, but for simplicity we assume the
electric field can be neglected in both frames.

%Sun7
One could include a time average in the definition of the mean flow,
equation \eqref{eq:meanu}.  \citet{penna2010} included a $100M$ time
average.  That is, they averaged the instantaneous flow over all of
$\phi$ and over $100M$ in time to obtain the mean flow.  At large
radii, $100M$ is much smaller than the orbital timescale, so this
extra averaging has little effect.  However, inside the ISCO, $100M$
is larger than the orbital timescale.  In this case, the time-averaged
mean flow tends to give larger $\alpha$.  The reason is that
time-averaging increases the discrepancy between the mean and
instantaneous flows by adding contributions to the mean flow from
earlier and later times.  This discrepancy propagates into $\alpha$
when the stress tensor is boosted to the rest frame of the mean flow.

\begin{figure}
\includegraphics[width=\columnwidth]{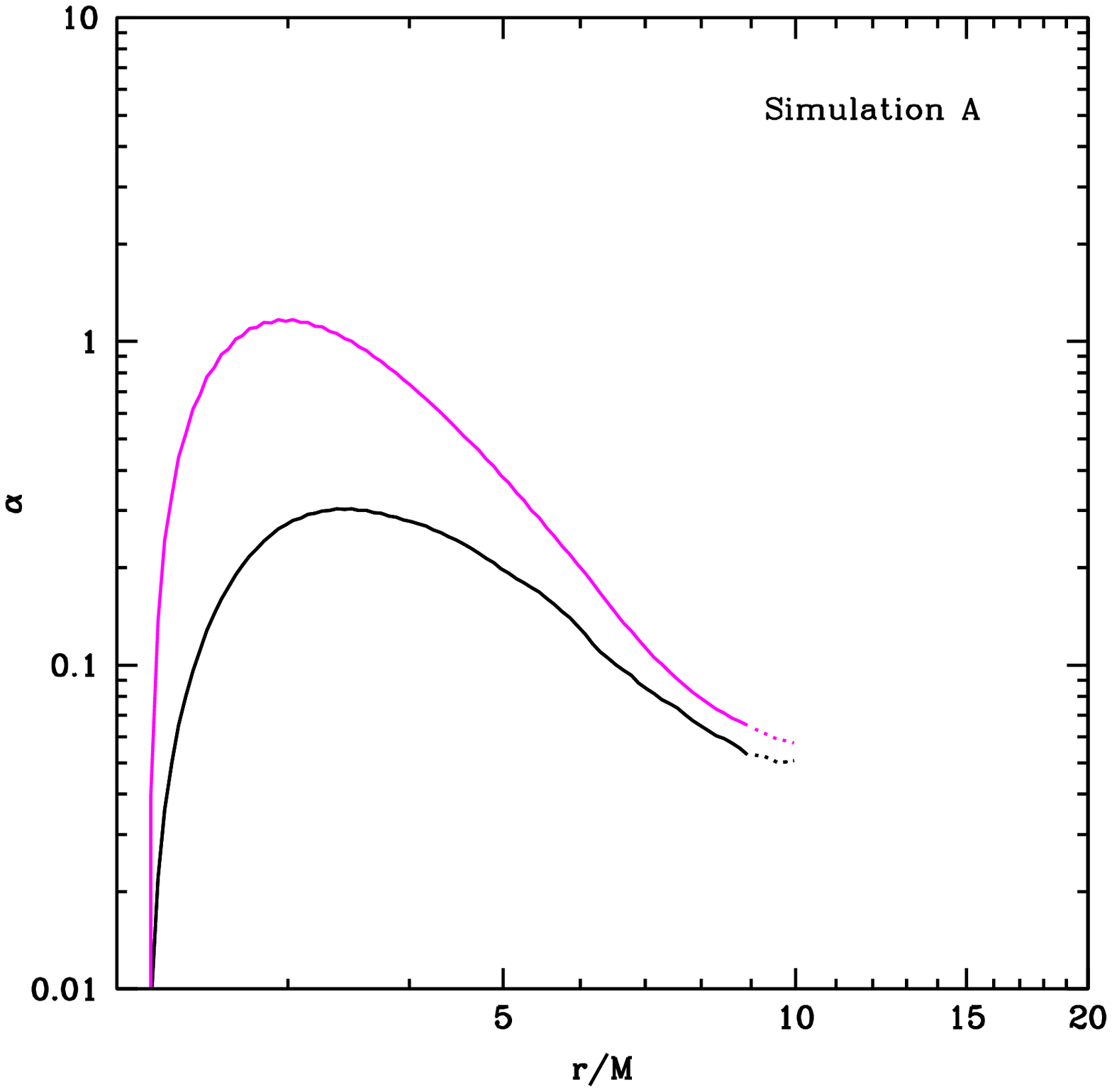}
\caption{Dimensionless viscosity parameter, $\alpha$, as a function of
  radius for simulation \thin, with (magenta) and without (black) a
  $100M$ time-average in the definition of the mean fluid frame,
  $\bar{u}^\mu$.  Outside the ISCO, $100M$ is longer than an orbital
  period, so the effect is small.  Inside the ISCO, $100M$ is several
  orbital periods, so the effect is large.}
\label{fig:afitavgs}
\end{figure}

%Sun8
Figure \ref{fig:afitavgs} shows $\alpha(r)$ for simulation \thin~with
and without a $100M$ time averaging in the definition of
$\bar{u}^\mu$.  Including the time averaging increases $\alpha$ and
the effect is greatest inside the ISCO.  In fact, the peak $\alpha$
exceeds unity when the mean flow is defined with a time average.  This
would imply the stresses carry more energy than the total energy of
the gas and magnetic fields, which is unphysical.  To avoid these
sorts of contradictions, we do not include any time averaging in
equation \eqref{eq:meanu}.

\begin{figure}
\includegraphics[width=\columnwidth]{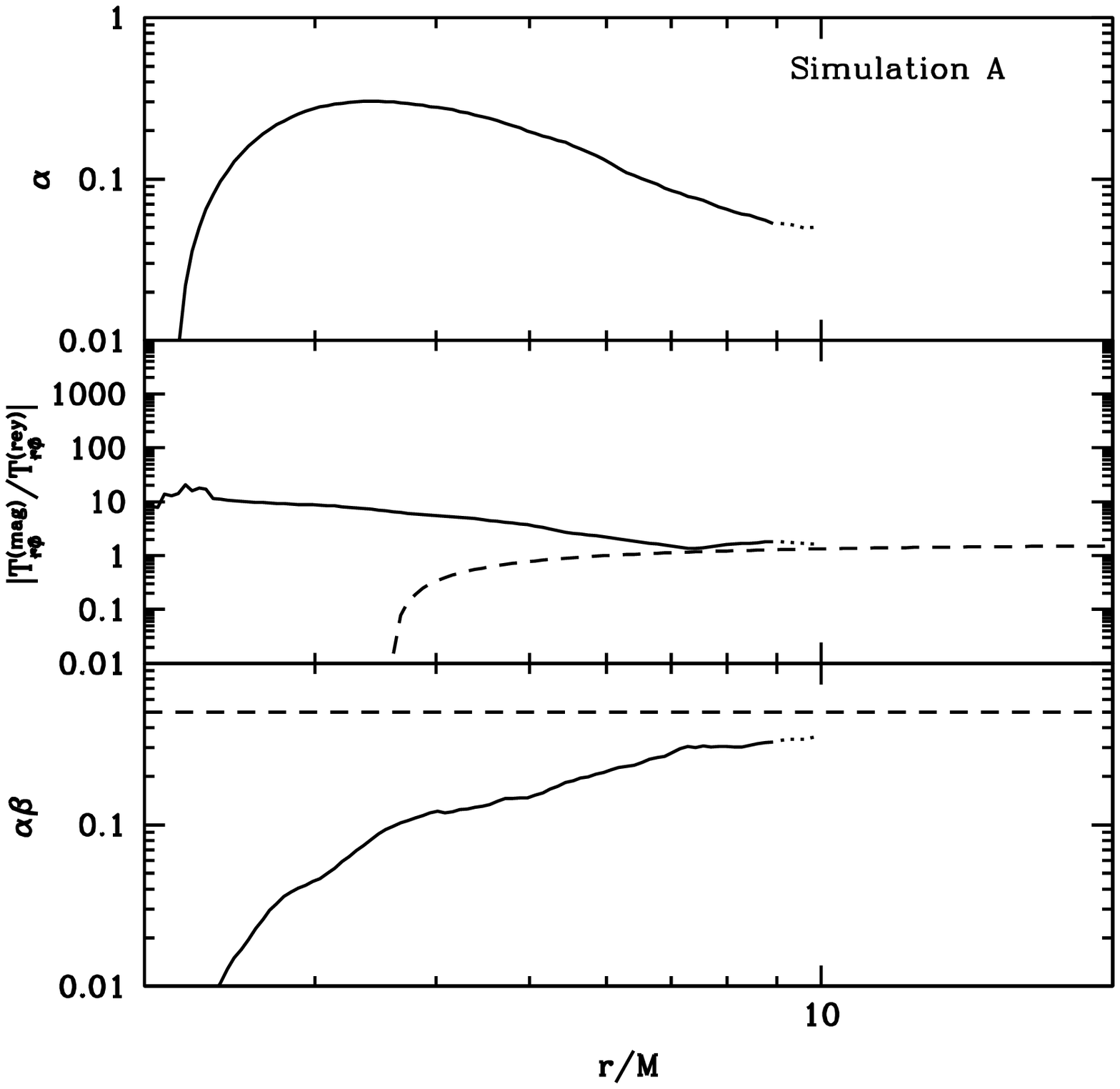}
\caption{\emph{Top panel:} Dimensionless viscosity parameter,
$\alpha$, as a function of radius for simulation \thin.  The data has
been time averaged from $t=7,000M$ to $20,000M$.  Solid and dotted
curves correspond to $r\leq \rsevere$ and $r\leq \rloose$,
\emph{Middle panel:} Time-averaged ratio of Maxwell to Reynolds stress
for simulation \thin~(solid and dotted curves). The dashed curve is
the the prediction from a linearized MRI analysis, $4/q(r)-1$
\citep{pessah2006}, for Keplerian $q(r)$, equation
\eqref{eq:qcirc}. \emph{Bottom panel:} Product $\alpha\beta$ for
simulation \thin~(solid and dotted curves).  The dashed line at
$\alpha\beta=0.5$ is the expected value for saturated MRI turbulence
\citep{blackman2008,guan2009,sorathia2012}.  The simulated product
falls below this value inside the ISCO where the flow is mostly
laminar.}
\label{fig:alphathin}
\end{figure}

%Sun9
The middle panel of Figure \ref{fig:alphathin} shows the ratio of
Maxwell stress to Reynolds stress in simulation \thin.  Outside the
ISCO, the ratio tracks the prediction of the linearized MRI, $(4-q)/q$
\citep{pessah2006}.  In fact, it is slightly larger, as is consistent
with shearing box simulations of MRI turbulence \citep{pessah2006}.
Inside the ISCO, the Maxwell stress is an order of magnitude larger
than the Reynolds stress.  The flow is mostly laminar inside the ISCO,
so Reynolds stress is weak, and the plunging fluid stretches the
magnetic field, so the Maxwell stress is strong.  The bottom panel of
Figure \ref{fig:alphathin} shows the product $\alpha\beta$.  It
approaches the expected value of $\approx 0.5$ in the disk
\citep{blackman2008,guan2009,sorathia2012}.

\begin{figure}
\includegraphics[width=\columnwidth]{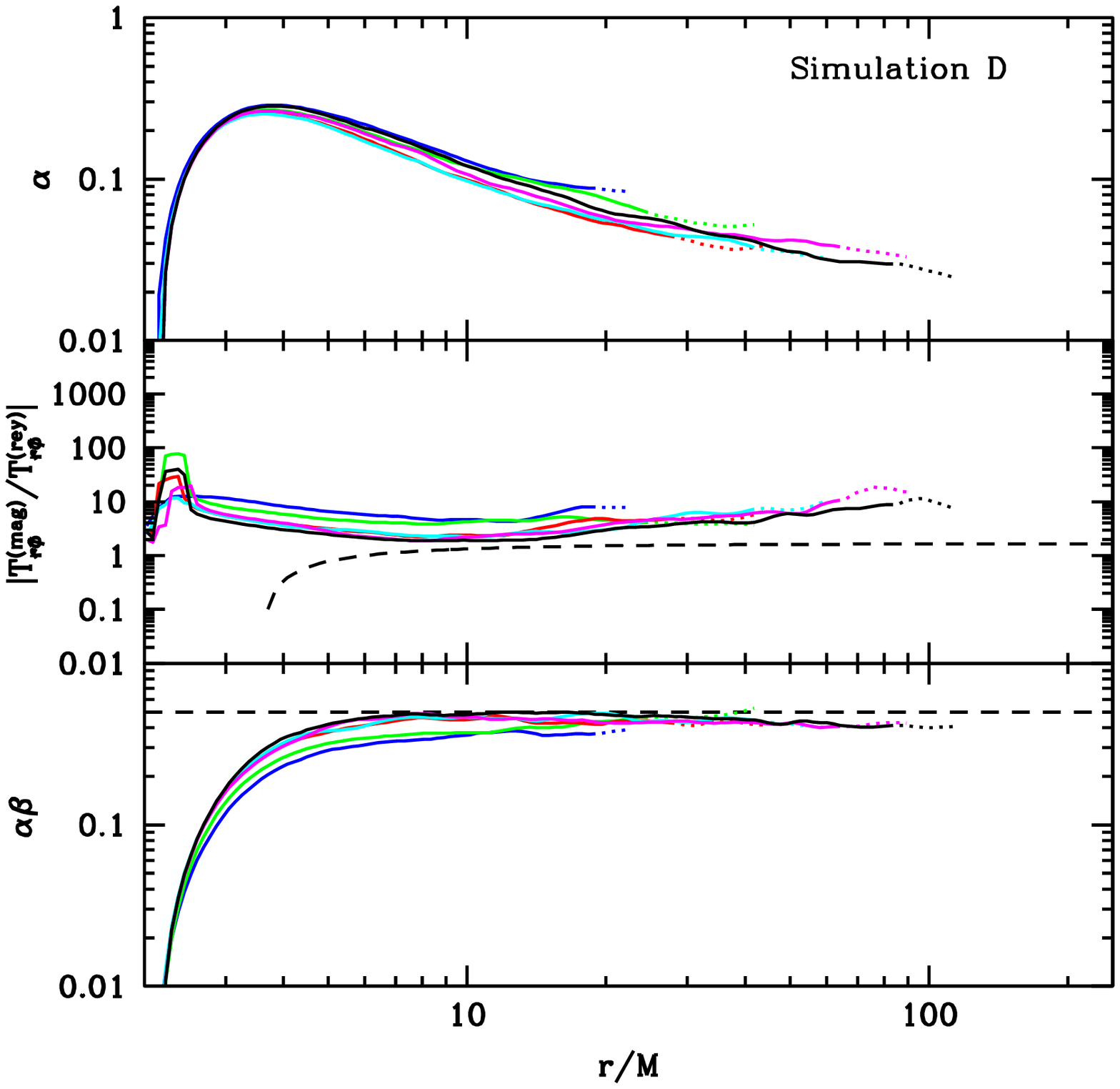}
\caption{Same as Figure \ref{fig:alphathin} but for simulation \sane.
Colors and line types are as in Figure \ref{fig:vrthin}.}
\label{fig:alphasane}
\end{figure}

%Sun10
The top panel of Figure \ref{fig:alphasane} shows $\alpha(r)$ for the
six time chunks of simulation \sane.  Outside the ISCO, the various
time chunks agree to within $30\%$, which provides an estimate for the
contribution of turbulent noise to the error in our $\alpha$
measurements.  Inside the ISCO, where the flow is more nearly laminar,
the agreement between the time chunks is better.

The bottom panel of Figure \ref{fig:alphasane} shows the ratio of
Maxwell to Reynolds stress as a function of radius for simulation
\sane.  Inside about $r=22M$, the ratio is consistent with the
ratio of simulation \thin.  Outside $r=22M$, the ratio begins
to grow with radius.  It is not clear what causes this.  It may
indicate that simulation \sane~has not reached quasi-steady state at these
radii yet.  Figure \ref{fig:alpha2dsane} shows highly magnetized,
irregular clumps of fluid in the disk region, even after
time-averaging the data over the last time chunk.  In a true
quasi-steady state, one would expect time averaging to eliminate these clumps.

\subsection{Comparison With the One-Dimensional $\alpha(r)$ Prescription}
\label{sec:simfits}

%Sun11
Finally, we compare $\alpha(r)$ of the simulations with the
one-dimensional prescription for $\alpha(r)$ of \S\ref{sec:1dalpha},
as defined by equation \eqref{eq:alphavsr} and the parameters listed
in Table \ref{tab:afits}.

\begin{figure}
\includegraphics[width=\columnwidth]{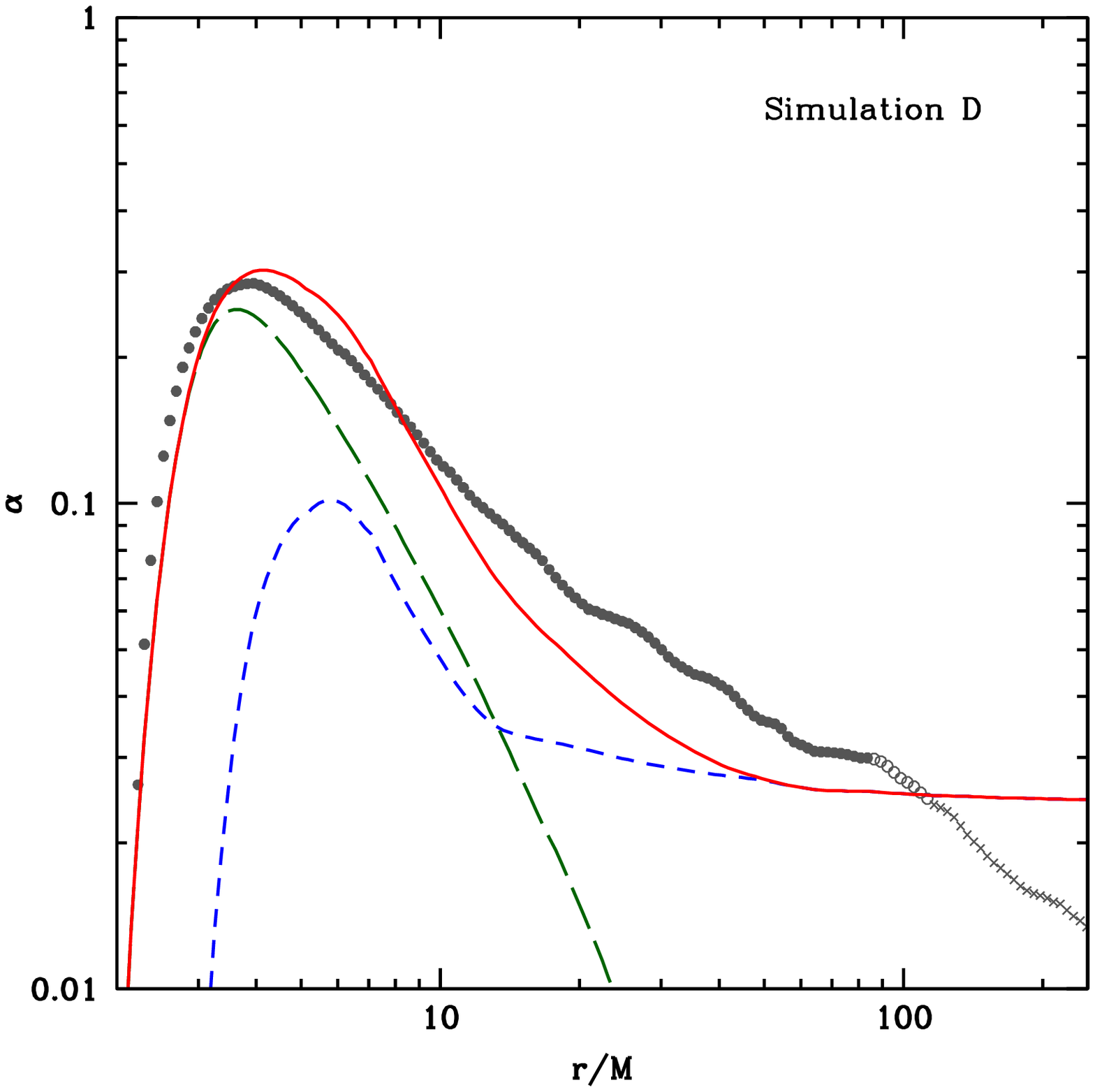}
\caption{Same as Figure \ref{fig:afitthin} but for simulation \sane.}
\label{fig:afitsane}
\end{figure}

Figure \ref{fig:afitthin} shows the agreement between the GRMHD
$\alpha(r)$ from simulation \thin~and the $\alpha(r)$ prescription.
Figure \ref{fig:afitsane} shows the agreement between simulation
\sane~and the $\alpha(r)$ prescription.  The interpretation of the
GRMHD $\alpha(r)$ in terms of a mean magnetic field component in the
inner regions, and a turbulent component in the outer regions, appears
to match the data.  It is more difficult to fit simulation
\sane~because the $\alpha(r)$ prescription relies on a sharp
distinction between the laminar, magnetically dominated inner regions
of the flow, and the turbulent, weakly magnetized outer regions.  This
distinction is cleanest in thin disks, like simulation \thin, which
have a clear inner edge at the ISCO.  In thick disks, like simulation
\sane, the inner edge of the disk is smeared out by the radial
velocity of the flow, so the separation of $\alpha$ into two
independent components is less sound.

The shape of $\alpha(r)$ does not depend strongly on all of the
parameters in Table \ref{tab:afits}.  The mass of the black hole in
the GRMHD simulations is dimensionless.  It is listed in Table
\ref{tab:afits} because it goes into the slim disk solutions that
underlie the one-dimensional $\alpha(r)$ prescription.  We set
$M/M_\odot=10$ arbitrarily and this choice has a negligible effect on
$\alpha(r)$.

The accretion rate also enters through the slim disk part of
$\alpha(r)$.  Our estimates of $\dot{M}/\dot{M}_{\rm edd}$ for the
GRMHD simulations are based on the analysis of \cite{zhu2012}, who
used $h/r$ as a proxy for the accretion rate.  This gives rough
estimates, which is all we need because the dependence of $\alpha(r)$
on the accretion rate is also very weak.

The magnetic flux threading the horizon, $\Upsilon$, is measured
directly from the GRMHD simulations.  It slightly affects the shape of
$\alpha(r)$ inside the ISCO.

The four parameters that strongly control the shape of $\alpha(r)$ are
$\alpha_0$, $\alpha_1$, $n$ and $\rgammie$.  It is encouraging that
$\alpha_0=0.025$, and $n=6$ give good fits to both simulations.  In
other words, both simulations have $\alpha(r)\sim 0.025$ at large
radii, and $\alpha \propto q^6$ (and $q>0$) in the turbulent disk.

Mean magnetic fields are only important inside the ISCO of simulation
\thin, so we set $\rgammie=6M$ in this case.  The region where mean
magnetic fields are important in simulation \sane~is broader, so we
set $\rgammie=30M$ in this case.

\begin{table*}
 \centering
 \begin{minipage}{70mm}
  \caption{Parameters for $\alpha(r)$ fits to the GRMHD simulations}
  \begin{tabular}{@{}ccccccccc@{}}
  \hline
  Simulation
  & $M/M_{\rm \odot}$
  & $a/M$
  & $\dot{M}/\dot{M}_{\rm edd}$
  & $\Upsilon$
  & $\alpha_0$
  & $\alpha_1$
  & $n$
  & $\rgammie$\\
  \hline
% 
% Sim.   & M  % a/M & Mdot & Ups   $ alph0  & alph1  & n & \rgammie
 \thin   & 10 & 0   & 0.5  & 0.6   & 0.025 & 100    & 6 & $r_{\rm ISCO}$ \\
 \thinA  & 10 & 0.7 & 0.2  & 3     & 0.025 & 10     & 6 & $r_{\rm ISCO}$ \\
 \thinLR & 10 & 0   & 0.2  & 6     & 0.025 & 1      & 6 & $r_{\rm ISCO}$ \\
 \sane   & 10 & 0   & 1    & 5     & 0.025 & 0.5    & 6 & 30M \\
 \saneA  & 10 & 0.7 & 1    & 10    & 0.025 & 0.5    & 6 & 30M \\
 \mad    & 10 & 0   & 1    & 30    & 0.025 & 0.1    & 6 & 30M \\
\hline \label{tab:afits}
\end{tabular}
\end{minipage}
\end{table*}

%435  
%2000 
%416  
%0.45 
%1    
%0.1  

\section{Analysis of Simulations \thinA, \thinLR, \saneA, and \mad}
\label{sec:moresims}

In this section we present data from four more GRMHD simulations.
This gives information about the dependence of the viscosity
parameters $\alpha_0$, $\alpha_1$, $n$, and $\rgammie$,
on black hole spin, resolution, and the amount of magnetic flux
threading the black hole.  Of these effects, the dependence on 
flux threading the black hole is the most dramatic.

\subsection{Simulation \saneA}

Simulation \saneA~is identical to simulation \sane~except the black
hole has spin parameter $a/M=0.7$ and the duration is $100,000M$.  We
have divided the simulation data into time chunks as we did for
simulation \sane, but there is one less time chunk because the
duration is half as long.  The time chunks and radii of convergence
estimates are listed in Table \ref{tab:sane7}.

\begin{table*}
 \centering
 \begin{minipage}{70mm}
  \caption{Time chunks for simulation \saneA}
  \begin{tabular}{@{}ccccc@{}}
  \hline
  Chunk & Time Range (M) & $\tchunk/M$ & $\rsevere/M$ & $\rloose/M$\\
 \hline
 I   & 3,000-6,000      & 3,000    & 9.5 & 15 \\ %blue
 II  & 6,000-12,000     & 6,000    & 15 & 20 \\ %green
 III & 12,000-25,000    & 13,000   & 22 & 31 \\ %red
 IV  & 25,000-50,000    & 25,000   & 31 & 44 \\ %cyan
 V   & 50,000-100,000   & 50,000   & 44 & 60 \\ %magenta
\hline \label{tab:sane7}
\end{tabular}
\end{minipage}
\end{table*}

\begin{figure}
\includegraphics[width=\columnwidth]{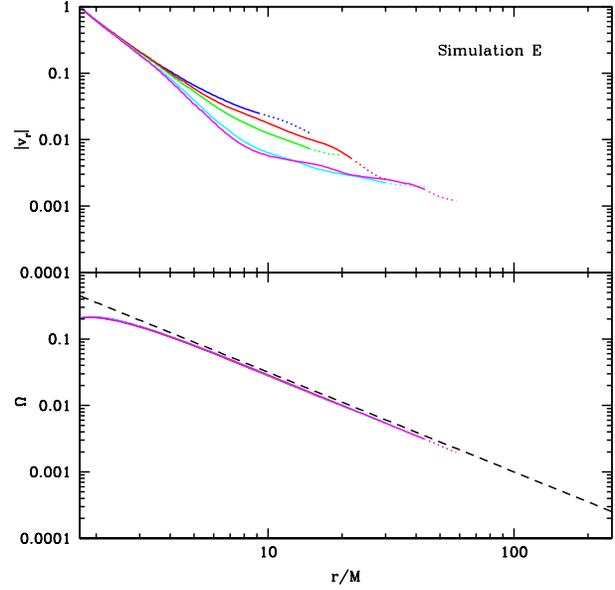}
\caption{Same as Figure \ref{fig:vrthin} but for simulation \saneA.}
\label{fig:vrsane7}
\end{figure}

Figure \ref{fig:vrsane7} shows the radial and angular velocities as a
function of radius for the five time chunks.  After the first three
time chunks, around $t=25,000$, the radial velocity drops by about a
factor of 2.  Then it holds steady (to within a few percent) for the
final two time chunks.  So even at $t=25,000M$, the simulation is
still settling down. The radial velocity at the end of simulation
\saneA~is about $30\%$ lower than the radial velocity at the end of
simulation \sane.  As a result, simulation \saneA~has converged over a
smaller range of radii; we find $\rloose=60M$ for the last time chunk
(simulation \sane~had $\rloose=90M$ over the same time interval.)

\begin{figure}
\includegraphics[width=\columnwidth]{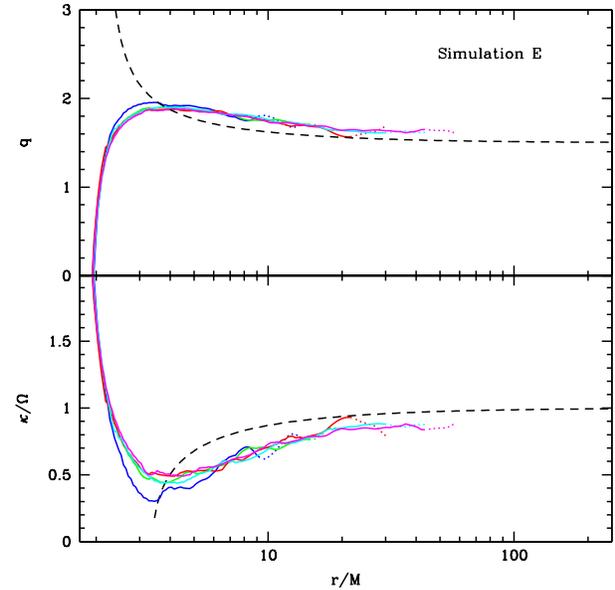}
\caption{Same as Figure \ref{fig:qthin} but for simulation \saneA.}
\label{fig:qsane7}
\end{figure}

Figure \ref{fig:qsane7} shows the dimensionless shear parameter and
epicyclic frequency as a function of radius.
Outside the ISCO, the shear parameter is about $20\%$ larger than the
shear parameter of circular geodesics and the epicyclic frequency is
about $20\%$ smaller, similar to the results from simulation \sane.
The minimum epicyclic frequencies of the two simulations are also
comparable. Simulation \sane~had $\kappa/\Omega \sim 0.6$ and simulation
\saneA~has $\kappa/\Omega \sim 0.5$.

\begin{figure}
\includegraphics[width=\columnwidth]{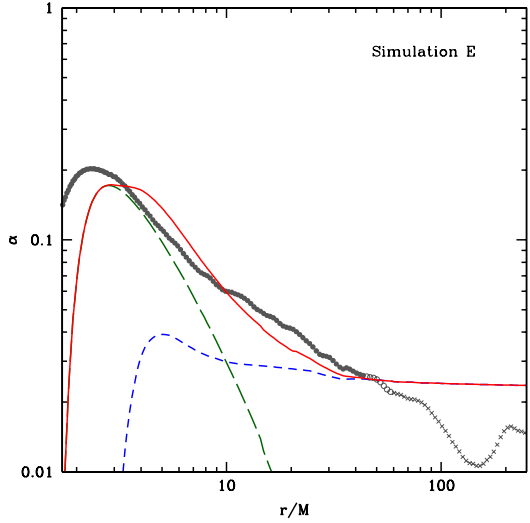}
\caption{Same as Figure \ref{fig:afitthin} but for simulation \saneA.}
\label{fig:afitsane7}
\end{figure}

Figure \ref{fig:afitsane7} compares $\alpha(r)$ as computed from the
last time chunk of the simulation against $\alpha(r)$ computed from
the one-dimensional viscosity prescription with $\alpha_0=0.025$,
$\alpha_1=0.5$, $n=6$, and $\rgammie=30M$.  This is the same choice of
parameters that gave a good fit to simulation \sane.
It is interesting that they fit simulation
\saneA~as well.  This suggests the parameters of the modified
viscosity prescription do not depend strongly on black hole spin.

\subsection{Simulation \thinA}

This simulation is identical to simulation \thin, except the black
hole is spinning with spin parameter $a/M=0.7$ and the resolution is
$256\times 64\times 32$ rather than $256\times 128\times 64$.

\begin{figure*}
\includegraphics[width=0.49\linewidth]{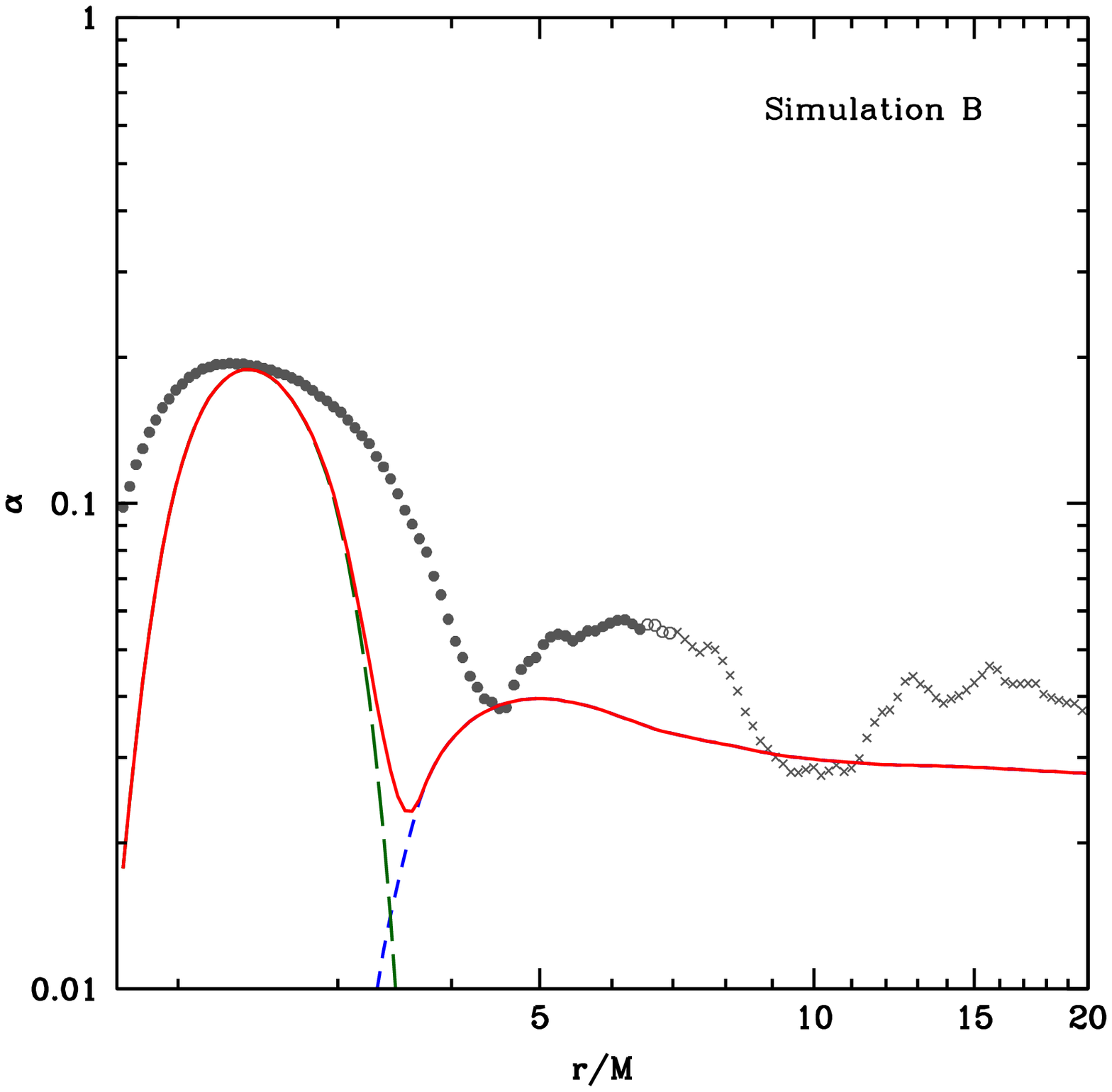}
\includegraphics[width=0.49\linewidth]{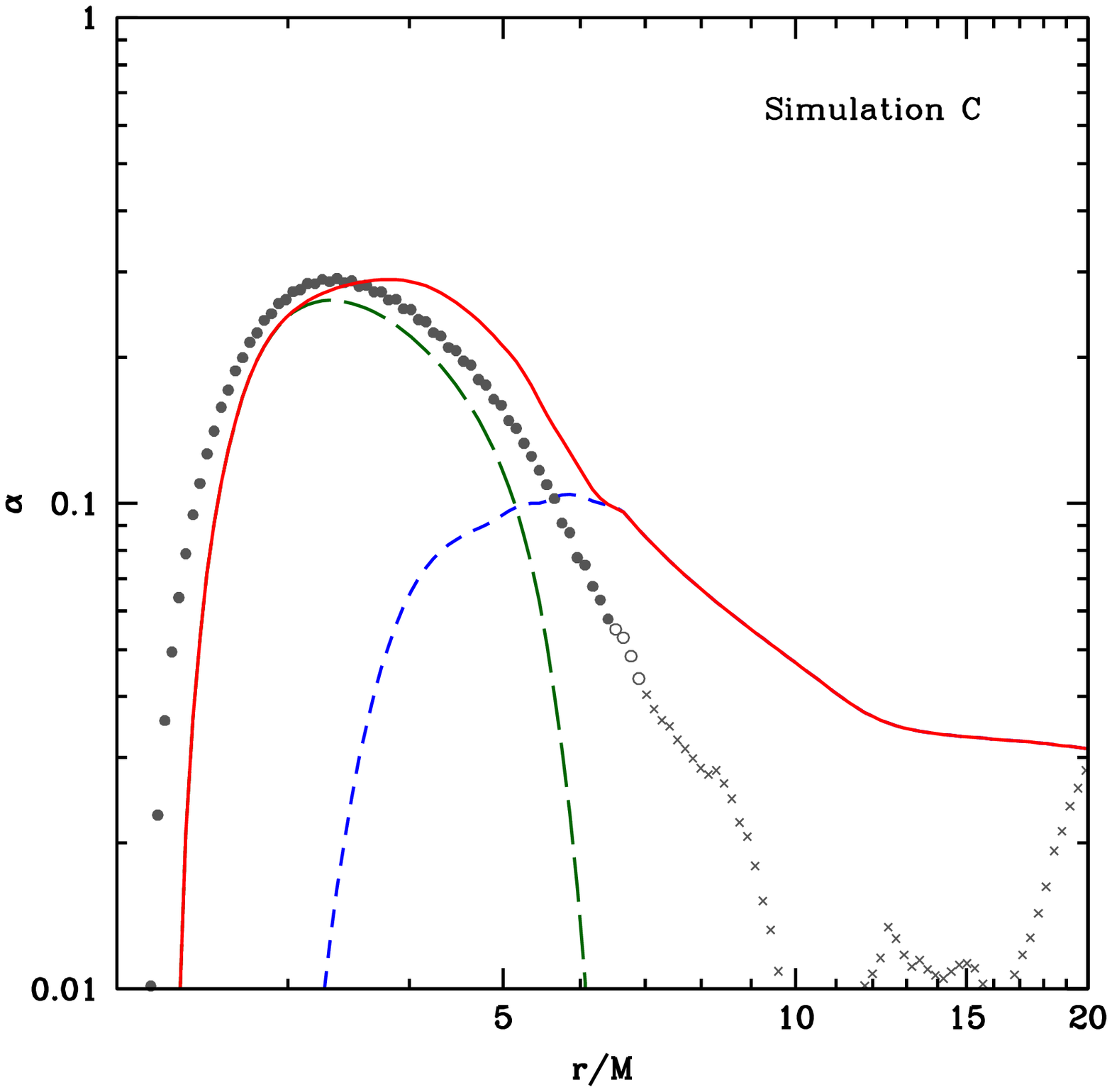}
\caption{Same as Figure \ref{fig:afitthin} except for simulations
  \thinA~(left panel) and \thinLR~(right panel).  }
\label{fig:afits}
\end{figure*}

The left panel of Figure \ref{fig:afits} shows the GRMHD $\alpha(r)$.
The $\alpha(r)$ prescription is shown for the same parameters that
gave a good fit to simulation \thin: $\alpha_0=0.025$, $\alpha_1=1$,
$n=6$, and $\rgammie=r_{\rm ISCO}$.  The parameters $\alpha_0=0.025$ and $n=6$
governing the turbulent part of $\alpha(r)$ are the same across all
four simulations we have considered so far, suggesting these
parameters do not depend strongly on disk thickness.

\subsection{Simulation \thinLR}

Simulation \thinLR~is the same as simulation \thin~except the
resolution is lower: $256\times64\times32$ versus $256\times128\times
64$ and the disk is thinner ($h/r\sim 0.05$ instead of $h/r\sim 0.1$).
The data from this simulation has a lower $\alpha$ than the
$\alpha(r)$ prescription with our fiducial choice of parameters
$\alpha_0=0.025$ and $n=6$.  
This suggests we are
under-resolving the MRI at this resolution.  To infer whether the
values $\alpha_0=0.025$ and $n=6$ are robust, higher
resolution simulations would be useful.

The results are shown in the right panel of Figure \ref{fig:afits}.

\subsection{Simulation \mad}

Simulation \mad~differs from simulation \sane~in one crucial respect.
The initial torus of gas is threaded with a single poloidal magnetic
field loop rather than multiple loops.  The center of the initial loop
is centered at $r=300M$ and gas from this radius does not reach the
black hole over the duration of the simulation.  So the polarity of
the flux that reaches the black hole is approximately constant and a
large net flux builds up on the hole.  \citet{narayan2012} give a
detailed account of the convergence in time and radius, and the role of
outflows, in simulations \sane~and \mad.

\begin{figure}
\includegraphics[width=\columnwidth]{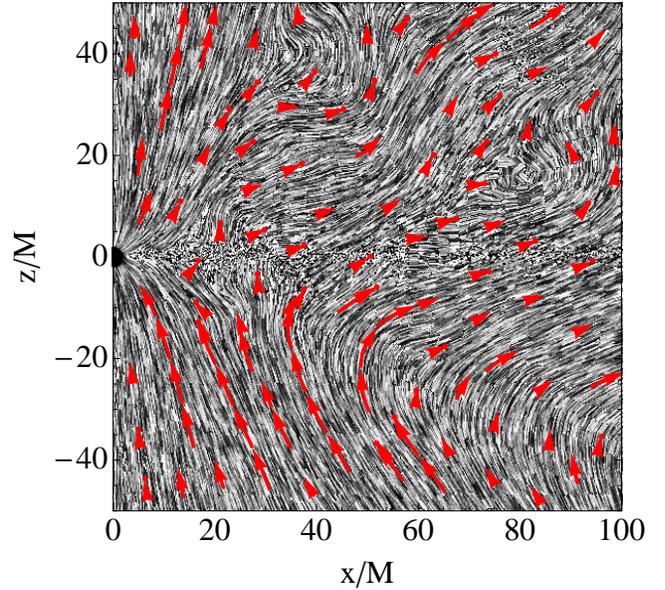}
\caption{Same as Figure \ref{fig:stream} but for Simulation \mad.
  The magnetic field structure does not show turbulent twisting on any
  scale; the flow is mostly laminar.
}
\label{fig:streammad}
\end{figure}

The large net flux carried by the gas in simulation \sane~has a
dramatic effect: the flow remains mostly laminar at all radii.
Figure \ref{fig:streammad} shows the fluid frame magnetic field in the
$r-\theta$ plane at $t=100,000M$, the final snapshot of the
simulation.  The eddies and turbulent twisting of the field are all but
gone on every scale, in marked contrast to the other five
simulations we considered (see, e.g., Figure \ref{fig:stream}).

\begin{figure}
\includegraphics[width=\columnwidth]{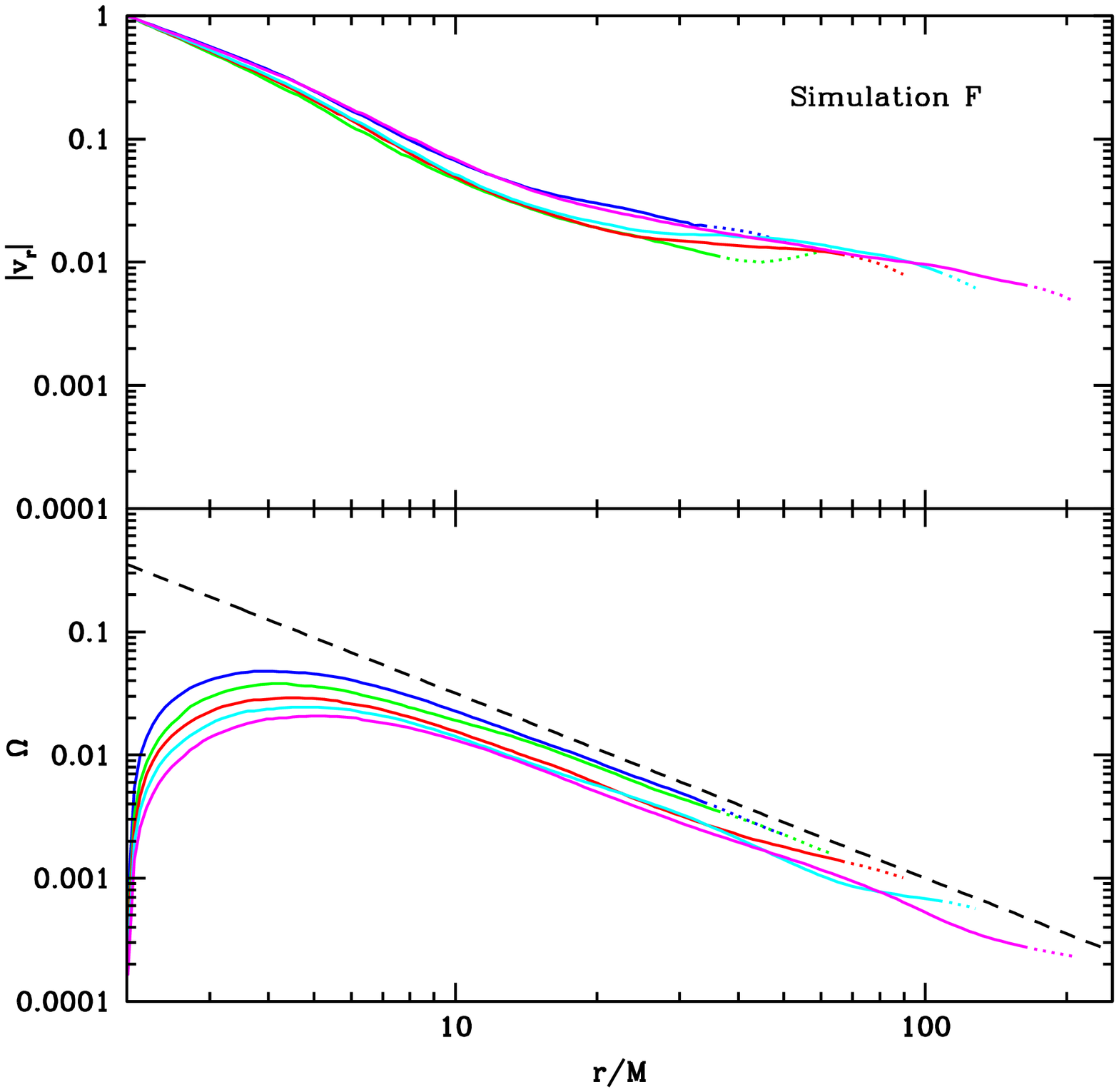}
\caption{Same as Figure \ref{fig:vrthin} but for simulation \mad.  Of
  the six simulations we consider, this simulation has the largest
  radial velocity and is the most sub-Keplerian.  }
\label{fig:vrmad}
\end{figure}

Following \citet{narayan2012}, we divide the simulation data into five
time chunks.  The time periods and estimated convergence radii for each
time chunk are summarized in
Table \ref{tab:mad}.  This simulation has the largest radial velocity
of any of the simulations (see Figure \ref{fig:vrmad}), so the
estimated convergence radii are the largest.  The final time chunk has
$\rsevere = 170M$ and $\rloose = 207M$.

Simulation \mad~also has the most sub-Keplerian angular velocity of
the six simulations (Figure \ref{fig:vrmad}).  The angular velocity
drops by a factor of a few between time chunks I and III, but it is
consistent across the final three time chunks to within a few percent.

\begin{table*}
 \centering
 \begin{minipage}{70mm}
  \caption{Time chunks for simulation \mad}
  \begin{tabular}{@{}ccccc@{}}
  \hline
   Chunk & Time Range (M) & $\tchunk/M$ & $\rsevere/M$ & $\rloose/M$\\
 \hline
 I   & 3,000-6,000      & 3,000    & 35 & 52 \\
 II  & 6,000-12,000     & 6,000    & 37 & 65 \\
 III & 12,000-25,000    & 13,000   & 69 & 90 \\
 IV  & 25,000-50,000    & 25,000   & 109 & 128 \\
 V   & 50,000-100,000   & 50,000   & 170 & 207 \\
\hline \label{tab:mad}
\end{tabular}
\end{minipage}
\end{table*}

\begin{figure}
\includegraphics[width=\columnwidth]{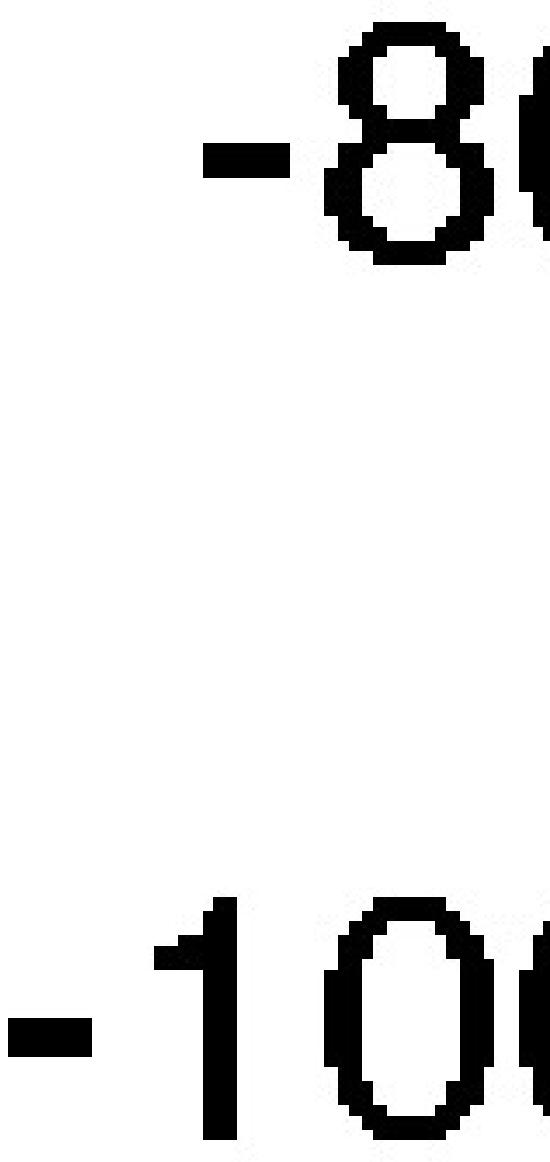}\\
\includegraphics[width=\columnwidth]{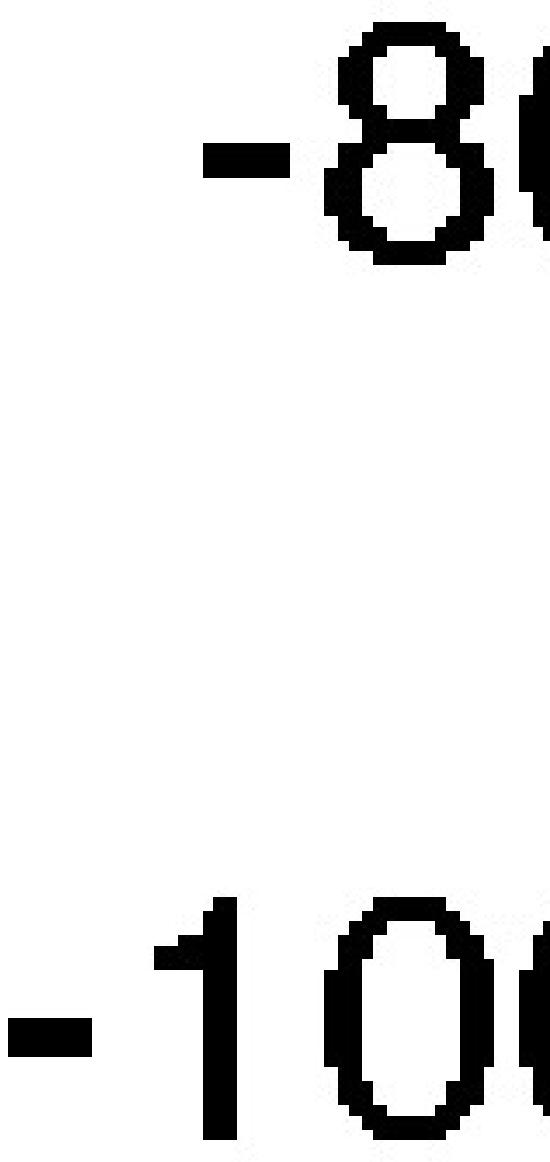}
\caption{Same as Figure \ref{fig:alpha2dthin} but for simulation \mad.
  A large, highly-magnetized, z-shaped clump persists over most of the
  flow inside $\rloose$.  The magnetized clump increases $\alpha$ and
  the ratio of Maxwell to Reynolds stress.  }
\label{fig:stressmad}
\end{figure}

The ratio of Maxwell stress to Reynolds stress has a much clumpier
distribution in the $r-\theta$ plane than any of the other
simulations.  A large, magnetized, z-shaped clump, where the Maxwell
stress is enhanced, extends throughout the flow (bottom panel of
Figure \ref{fig:stressmad}).  The irregular shape of the clump
suggests it is a non-equilibrium structure.  Perhaps if the duration
of the simulation was longer it would be smoothed out.  The appearance
of the magnetized clump suggests $\rsevere = 170M$ is a better estimate
for the radii of convergence for this simulation than $\rloose=207$.

\begin{figure}
\includegraphics[width=\columnwidth]{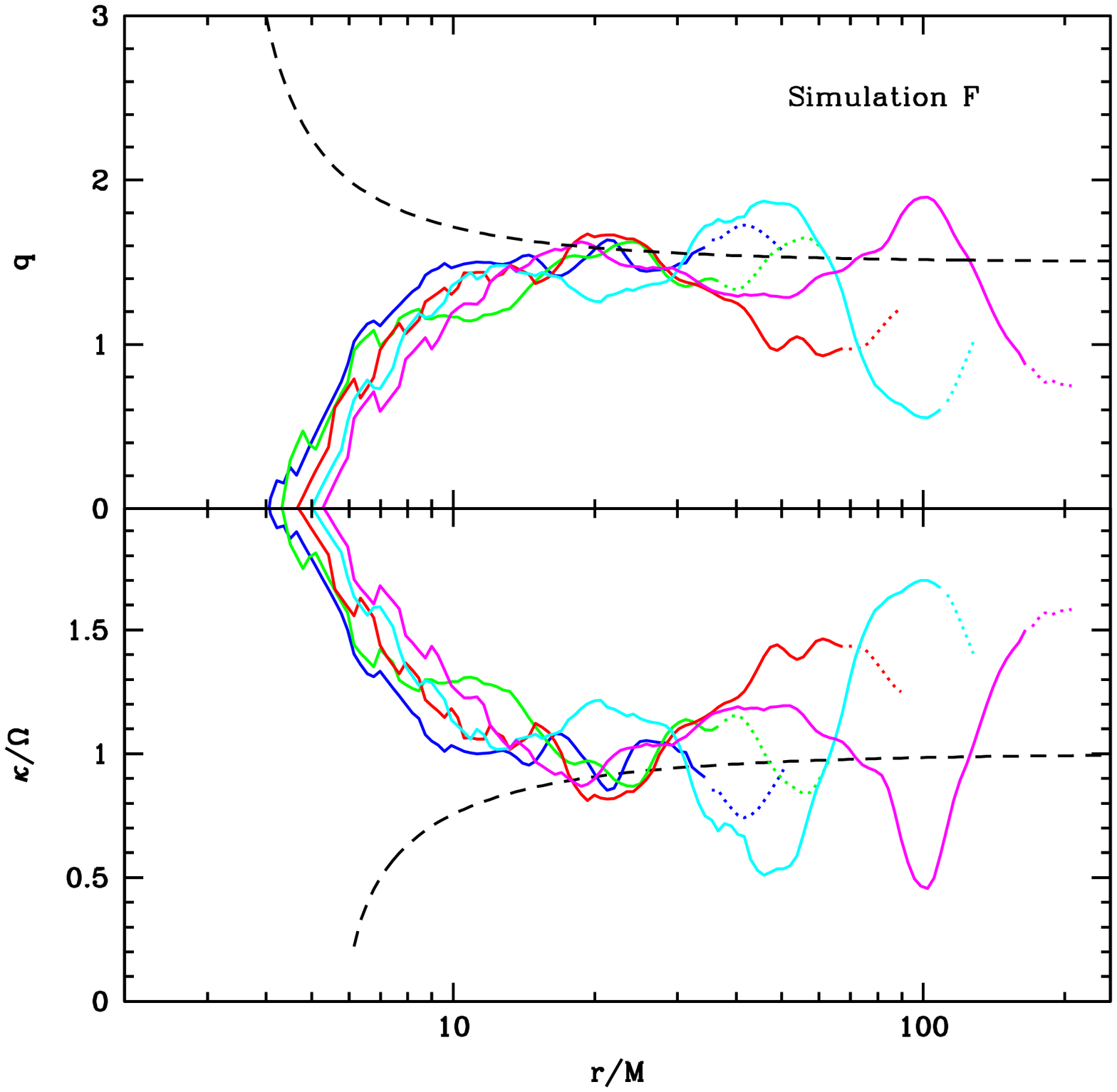}
\caption{Same as Figure \ref{fig:qthin} but for simulation \mad.
}
\label{fig:qmad}
\end{figure}

The shear parameter of the flow (top panel of Figure \ref{fig:qmad})
varies by about $50\%$ between time chunks.  Outside the ISCO, the
shear parameter is roughly constant with radius.  The epicyclic
frequency (bottom panel of Figure \ref{fig:qmad}) has its minimum near
$r=20M$, rather than at the ISCO.  The minimum itself is very broad
and shallow, not extending much below $\kappa/\Omega=1$.  In other
words, the inner edge of the disk has moved well outside the ISCO and
is highly smeared out.

\begin{figure}
\includegraphics[width=\columnwidth]{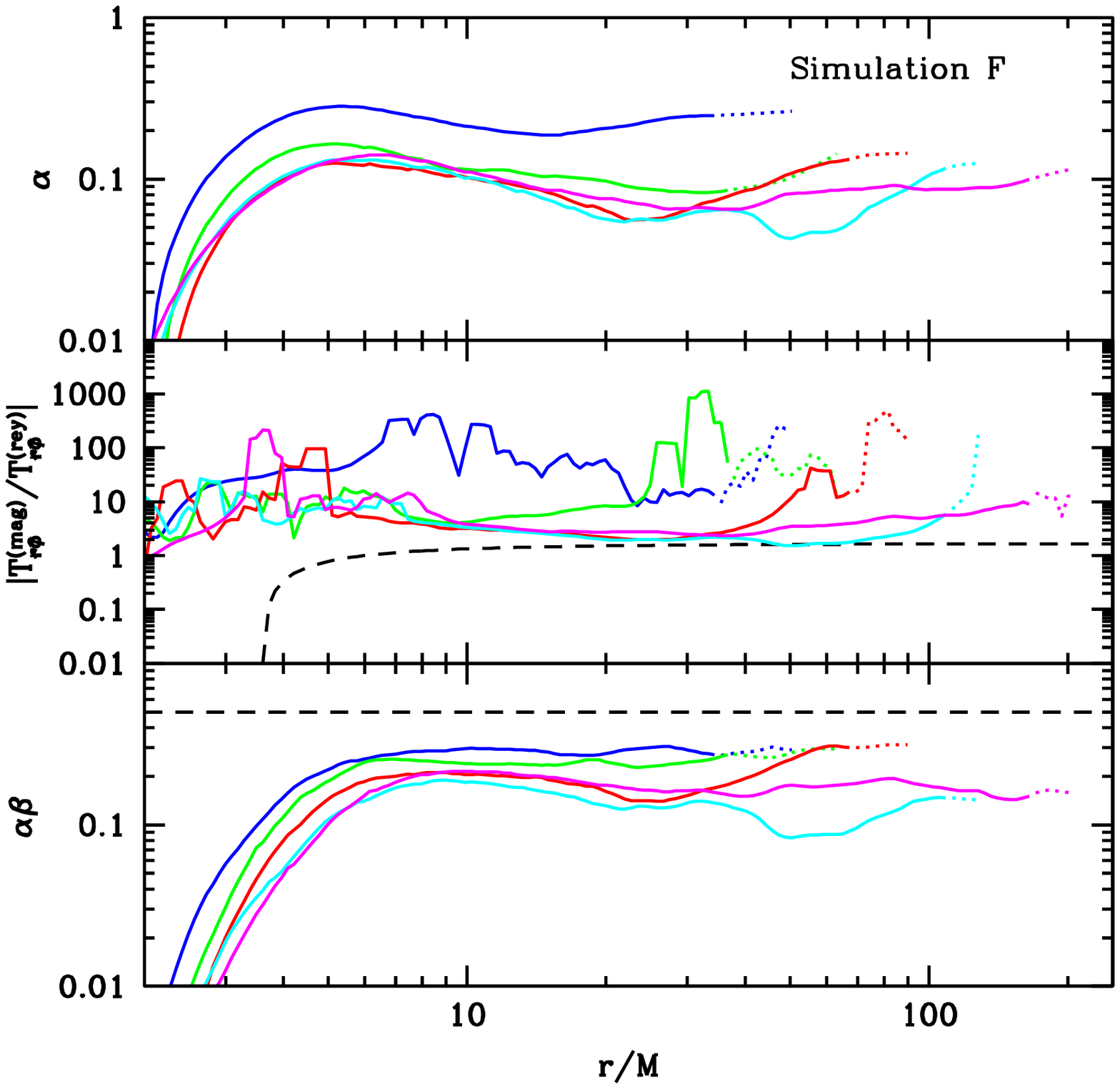}
\caption{Same as Figure \ref{fig:alphathin} but for simulation \mad.}
\label{fig:alphamad}
\end{figure}

The profiles of $\alpha$ as a function of radius for the five time
chunks are shown in the top panel of Figure \ref{fig:alphamad}.
Outside $r\approx 20M$, the profiles of $\alpha$ are constant with
radius, even increasing slightly.  The other simulations have $\alpha$
decreasing with radius.  This suggests the turbulent contribution to
$\alpha$ is not dominating even at the largest converged radii, which
is consistent with the laminar structure of the magnetic field lines.

The bottom panel of Figure \ref{fig:alphamad} shows the ratio of
Maxwell to Reynolds stress as a function of radius for the five time
chunks.  For the first two time chunks, the Maxwell stress is an order
of magnitude larger than the Reynolds stress at all radii.  During the
final three time chunks, the ratio appears to have stabilized out to
$r=40M$.  At larger radii, the Maxwell stress is enhanced by the
non-equilibrium, magnetized, z-shaped clump noted earlier.

\begin{figure}
\includegraphics[width=\columnwidth]{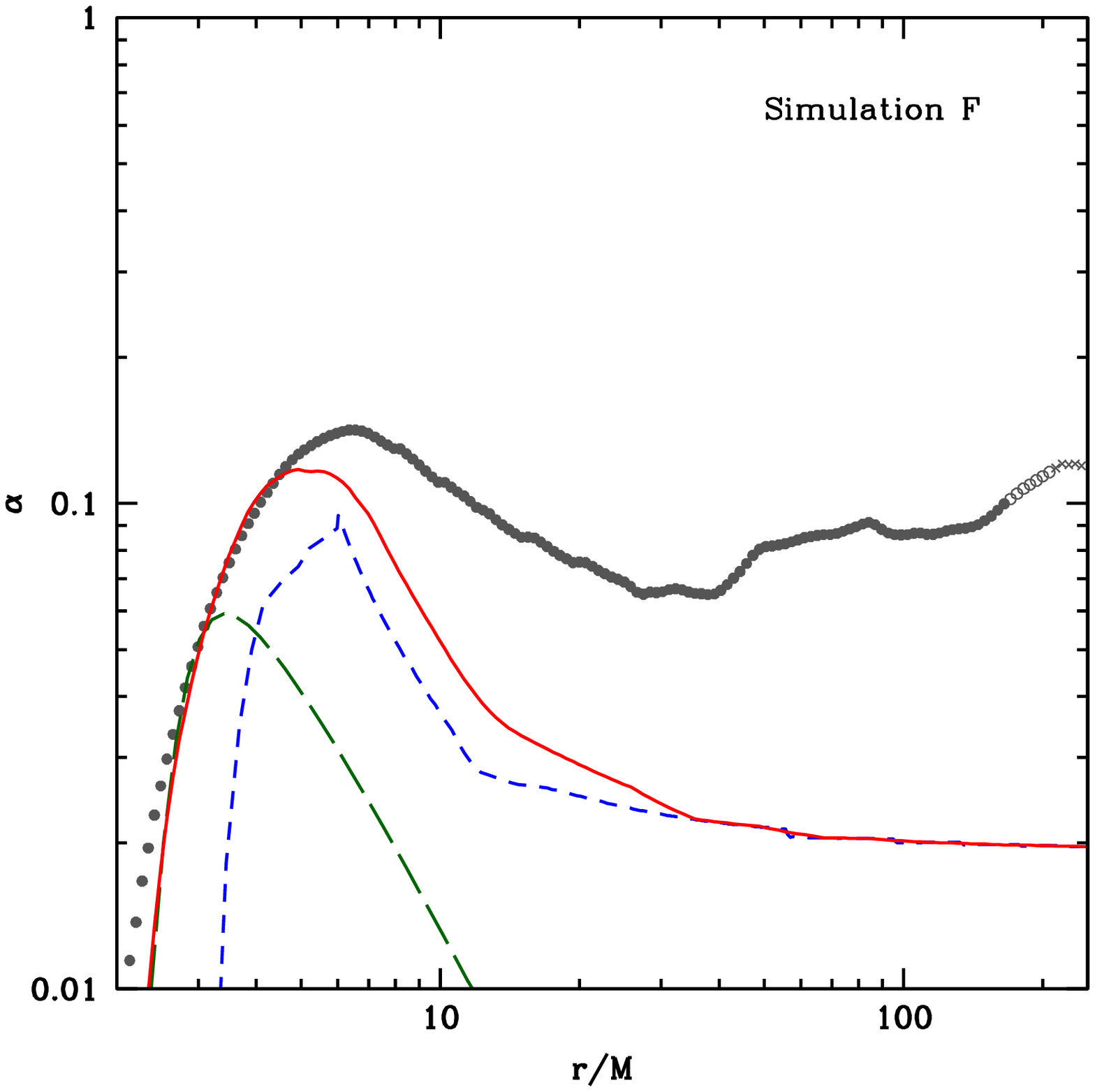}
\caption{The dimensionless viscosity parameter, $\alpha$, of
  simulation \mad~as a function of radius.  We have been unable to
  find a good fit to the simulation data using our $\alpha(r)$
  prescription.}
\label{fig:afitmad}
\end{figure}

Our $\alpha(r)$ prescription does not appear to give a good fit to the
simulation \mad~results (Figure \ref{fig:afitmad}).
This is probably because the simulation is mostly laminar at all radii
(as shown by Figure \ref{fig:streammad}), whereas our $\alpha(r)$
prescription assumes turbulence dominates the stress beyond the
innermost radii.  For simulations \thin--\saneA, this is a good
assumption provided the disk region of the flow is distinguished from
the coronal regions.  However, the entire domain of simulation \mad~is
mostly laminar and highly magnetized, and so it should perhaps be
considered entirely coronal gas.  It appears a different $\alpha(r)$
prescription is needed to describe such highly magnetized flows.

\section{$\alpha$-Disk Solutions with Variable $\alpha(\lowercase{r})$}
\label{sec:slim}

In this section, we evaluate the dependence of $\alpha$-disk solutions
on the $\alpha(r)$ prescription.  The particular $\alpha$-disks
we consider are ``slim disks'' \citep{abramowicz1988,sadowski2011}.  We
compare slim disk solutions with constant $\alpha=\alpha_0$ to
solutions with varying $\alpha(r)$, where $\alpha(r)$ is defined by
equation \eqref{eq:alphavsr} and the parameters inferred from
simulation \thin~(c.f. Table \ref{tab:afits}).  That is, we consider a
non-spinning, 10 solar mass black hole, threaded with a magnetic flux
$\Upsilon=0.6$.  The viscosity parameters are $\alpha_0=0.025$,
$n=6$, $\alpha_1=100$, and $\rgammie=6M$.  We consider two different
accretion rates: $30\%$ Eddington and Eddington.

\begin{figure}
\includegraphics[width=\columnwidth]{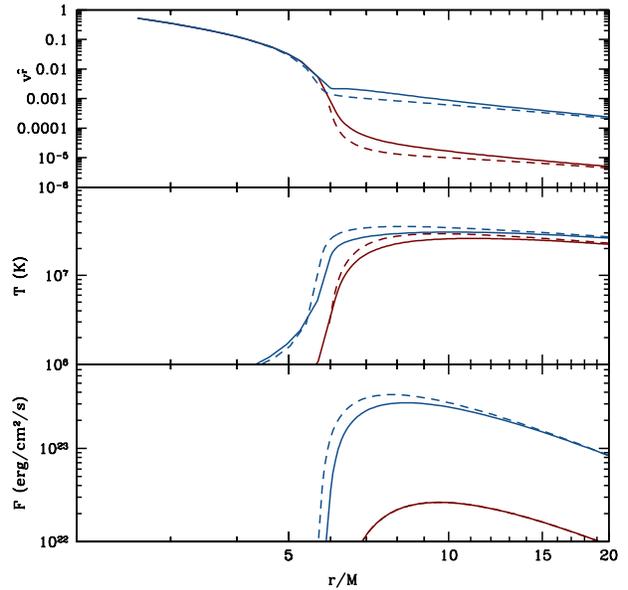}
\caption{Slim disk solutions with varying $\alpha(r)$ (solid curves)
  and with constant $\alpha=0.025$ (dashed curves), for the parameters
  inferred from simulation \thin~(c.f. \ref{tab:afits}).  Solutions
  are shown at the Eddington accretion rate (blue curves) and at
  $30\%$ Eddington (red curves).
\emph{Top panel:} Radial velocity as a function of radius.  Solutions
with varying $\alpha(r)$ have larger radial velocities.
\emph{Middle panel:} Midplane temperature (not effective temperature)
as a function radius.  Solutions with varying $\alpha(r)$ are colder
in the X-ray emitting portions of the flow.
\emph{Bottom panel:} Radiant flux as a function of radius.  Radiant
flux is affected by $\alpha(r)$ only at high accretion rates.
}
\label{fig:vfitthin}
\end{figure}

Figure \ref{fig:vfitthin} shows our results.  At large radii,
$\alpha(r)$ converges to $\alpha_0$, so the solutions with constant
and varying $\alpha(r)$ are the same to within a percent.  Inside the
ISCO, the fluid plunges toward the black hole with little dissipation,
so in the innermost regions the solutions are again insensitive to the
$\alpha$ prescription.  Only in an intermediate zone, between the ISCO
and $r\approx 20M$, does the shape of $\alpha(r)$ have a significant
effect.  For solar mass black holes, this region emits predominately
in X-rays and is relevant for black hole spin measurements
\citep{gou2011}.

In this zone, the $\alpha(r)$ prescription has a larger $\alpha$ than
the constant $\alpha=\alpha_0$ prescription.  So the $\alpha(r)$
prescription increases the disk's radial velocity by a factor of $2-3$
and lowers its central (mid-plane) temperature by about $50\%$.  In
fact, a solution accreting at the Eddington limit with varying
$\alpha(r)$ has the same central temperature as a solution accreting
at $30\%$ Eddington with constant $\alpha$ (c.f. Figure
\ref{fig:vfitthin}).  So the $\alpha(r)$ prescription has a
significant effect on central temperature.  Central temperature
depends on both effective temperature and optical depth, so the effect
is really due to changes in surface density.

At low accretion rates, the disk is radiatively efficient and the
effect of $\alpha(r)$ on the radiated fluxed is negligible.  At high
accretion rates, advection becomes important and the radiated flux
shows its dependence on $\alpha(r)$.  At the Eddington limit, the flux
from the solution with varying $\alpha(r)$ is about $50\%$ lower than
the flux from the solution with constant $\alpha$.  So flux is only
affected by the $\alpha(r)$ prescription at high accretion rates.

\section{Discussion and summary}
\label{sec:discuss}

\begin{figure}
\includegraphics[width=\columnwidth]{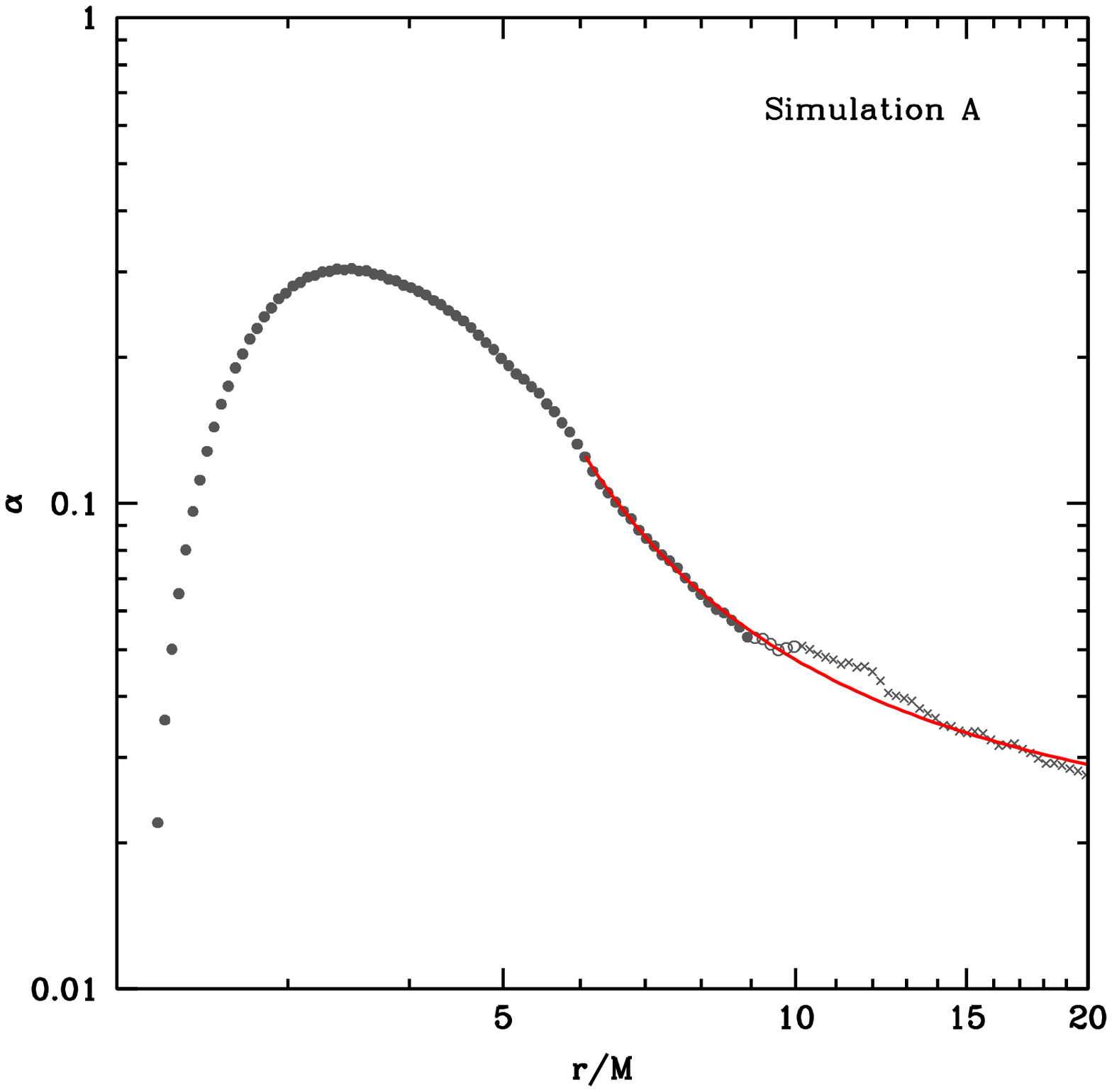}
\caption{
Analytical $\alpha(r)$ prescription
defined by equation \eqref{eq:alphasimple} for $a/M=0$ (solid red
curve).  This model gives a good fit to data from Simulation \thin.
Data from inside $\rsevere$ are marked
with filled circles, data from between $\rsevere$ and $\rloose$ are
marked with open circles, and data from outside $\rloose$ are marked
with crosses (c.f. \S\ref{sec:simalpha}).}
\label{fig:afitsimple}
\end{figure}

The $\alpha(r)$ prescription of \S\ref{sec:1dalpha} must be computed
numerically.  However, for all practical purposes, the function
$\alpha(r)$ for thin disks can be reduced to a simple analytical
formula.  Our simulations suggest
\begin{equation}\label{eq:alphasimple}
\alpha(r) = 0.025 \left[\frac{q(r)}{3/2}\right]^6, \quad (q>0),
\end{equation}
where $q(r)$ is given analytically by equation \eqref{eq:qcirc}.  The
constant coefficient and exponent in equation \eqref{eq:alphasimple}
are the values favored by our GRMHD simulations (c.f. Table
\ref{tab:afits}).  They will change as better simulation data becomes
available.  The free parameters are the mass and spin of the black
hole, which enter through equation \eqref{eq:qcirc} for $q(r)$.  The
contribution from mean magnetic fields can be ignored in this
approximation because mean field stresses are only significant inside
the ISCO, and thin disks are not sensitive to $\alpha(r)$ in this
region.  Equations \eqref{eq:qcirc} and \eqref{eq:alphasimple} thus
give an analytical $\alpha(r)$ prescription that can be used for thin
disk models (see Figure \ref{fig:afitsimple}).  The more general $\alpha(r)$ prescription of
\S\ref{sec:1dalpha} is needed for thick disks.

To summarize our main results, we have constructed a one-dimensional
prescription for $\alpha(r)$ and estimated parameters for this
prescription based on data from GRMHD simulations.  The fact that $\alpha$
varies with radius had been anticipated long ago
\citep{pringlerees1972,SS73} but global MHD simulations provide the first
quantitative information about the shape of $\alpha(r)$
\citep{papaloizou2003,penna2010,penna2012,fromang2011}.

Our modified $\alpha(r)$ prescription, equation \eqref{eq:alphavsr},
is the sum of two components.  The first component describes
mean field stresses.  It is important in the laminar, inner regions
of accretion disks, where the plunging fluid stretches the frozen-in
field.  Our description of this component is based on the model of
\citet{gammie1999}, which supplies the magnetic stress as a
function of radius and two free parameters: black hole spin and the amount
of magnetic flux threading the horizon.  

The second component of the $\alpha(r)$ prescription describes
turbulent stresses.  As emphasized by \citet{pessah2006}, $\alpha$
depends on the shear parameter, $q$.  In Newtonian gravity, Keplerian
flow has a constant shear parameter, $q=3/2$, but general relativistic
corrections give even Keplerian disks around black holes a varying
$q(r)$, as discussed in \S\ref{sec:prelims}.  The shear parameter
increases from $q=3/2$ at the outer edge of the disk to $q=2$ at the
inner edge of the disk.  This is a $50\%$ change in $q$ but it creates
a larger variation in $\alpha$, because $\alpha$ goes as $q^n$ (for $q>0$).  Our
GRMHD simulations are too noisy to infer $n$ precisely, but the data
seem to prefer $n\approx 6$.  This is consistent with the simulations
of \citet{pessah2006}, which resulted in $n$ between 2 and 8.
Analytical MHD closure models for the MRI also allow $n$ between 2 and 8
\citep{kato1993,kato1995,ogilvie2003,pessah2006b,pessah2006,pessah08}.

Simple extensions of the standard $\alpha$ prescription and some
closure models predict negative $\alpha$ for $q<0$
\citep{kato1993,kato1995,ogilvie2003}.  An exception is the closure
model of \citet{pessah2006b,pessah2006}.  Data from shearing box
simulations show zero turbulent stress for $q<0$ \citep{pessah08}.
Our simulations are consistent with this data, although they are not
as decisive on this point because negative shear parameters are only
found inside roughly the photon orbit, where mean field stresses are
large.  Our prescription for $\alpha(r)$ is always positive and the
turbulent contribution vanishes for $q<0$.  The mean field term in our
$\alpha(r)$ prescription gives a good description of our simulation
data in regions near the black hole where $q<0$.

We have discussed accretion onto black holes.  When a disk accretes onto a star,
a boundary layer forms between the star and disk.  It can generate
half the accretion luminosity in soft X-rays \citep{pringle1977}.  The
boundary layer in stellar accretion is similar to the region inside
the ISCO in black hole accretion.  In both regions, the angular
velocity is non-Keplerian and the shear amplifies the magnetic field
\citep{armitage2002,steinacker2002}.  \citet{steinacker2002} found
$\alpha(r)$ profiles in MHD simulations of boundary layers that are
similar to our $\alpha(r)$ profiles inside the ISCO.  It would be
interesting to extend the $\alpha(r)$ prescription to these cases.

We analyzed data from six GRMHD accretion disk simulations.
Three of the simulations are thin, radiatively efficient accretion
disks (simulations \thin, \thinA, and \thinLR).  The other three are
geometrically thick, radiatively inefficient accretion flows
(simulations \sane, \saneA, and \mad).  The simulations vary in
resolution from $256\times 64\times 32$ to $256\times 128\times
64$.  Two of the simulations describe spinning black holes, with spin
parameter $a/M=0.7$ (simulations \thinA~and \saneA) and the others
describe non-spinning black holes.  MRI driven turbulence and large
scale magnetic fields generate stresses in the simulated disks
self-consistently, so the $\alpha$ viscosity prescription is not
assumed.  Instead, we measure $\alpha(r)$ from the simulation data.

For each simulation, we measure $\alpha(r)=T_{\hr\hp}/p_{\rm tot}$
and compare it against the $\alpha(r)$ prescription.  Thin accretion
disks are easier to describe with the $\alpha(r)$ prescription than thick accretion
disks because they have a sharp transition at the ISCO that creates
two distinct regions: a magnetically dominated region inside the ISCO,
and a weakly magnetized, turbulent region outside the ISCO.  This
distinction is blurred in thick accretion disks by the large radial
velocity of the flow.  So some of the simplifications in the $\alpha(r)$
prescription are less applicable.  Simulation \mad, in which a large
magnetic flux was allowed to build up on the hole, is particularly
difficult to interpret because turbulence is almost completely absent.

We were careful to only include fluid which has reached a quasi-steady
state.  The timescale to reach quasi-steady state scales as the
inverse of radial velocity, and is thus an increasing function of
radius.  So the inner regions of disks converge before the outer
regions.  Thin accretion disks have smaller radial velocities than
thick accretion disks, so thin disk simulations are only converged out
to $r\approx 10M$ rather than $r\approx 100M$.

For further insight into the simulations, we analyzed their shear
parameters and epicyclic frequencies.  Outside the ISCO, the shear
parameters of the simulations are usually within $20\%$ of the
Keplerian prediction.  Inside the ISCO, the shear parameter turns over,
going to zero near the photon orbit at $r=3M$ where the angular
velocity peaks.  So the turbulent contribution to $\alpha$, which scales
as $q^n$ (for $q>0$), is unimportant near the black hole.

The epicyclic frequency of the flow is also close to the Keplerian
value outside the ISCO.  The inner edge of the flow can be identified
with the minimum in $\kappa(r)$, the radius where the flow is most
unstable.  This is usually near the ISCO, although when the disk is
thick the inner edge is smeared out by the large radial velocity of
the flow.

Finally, we considered the effect of the $\alpha(r)$ prescription on
$\alpha$-disk solutions.  We compared solutions with varying
$\alpha(r)$ to solutions with constant $\alpha=\alpha_0$.  We fixed
the free parameters $\alpha_0, \alpha_1, n$, and $\rgammie$ using the
values inferred from simulation \thin~(c.f. Table \ref{tab:afits}).
The differences between varying and constant $\alpha$ are only
significant in the region between the ISCO and $r\approx 20M$.  At
smaller radii, the gas is plunging too quickly for stresses to act, so
$\alpha$ does not enter, and at larger radii the varying $\alpha(r)$
prescription is converging to $\alpha=\alpha_0$.  In the intermediate
zone between the ISCO and $r\approx 20M$, the solutions with varying
$\alpha(r)$ have larger $\alpha$ than the solutions with constant
$\alpha$.  This increases their radial velocity and lowers their
central temperature and radiant flux.  The effect on central
temperature is most significant.  A solution accreting at the
Eddington rate with radially varying $\alpha(r)$ has the same central
temperature as a solution accreting at $30\%$ Eddington with constant
$\alpha$.  Central (mid-plane) temperature depends on both the
effective temperature and the optical depth, so the effect is really
due to changes in surface density. The effect of $\alpha$ on the
radiant flux is only important at high accretion rates.

The main shortcomings of our $\alpha(r)$ prescription are the absence
of gas pressure in the prescription for the mean magnetic field
component, and the absence of magnetic fields in the prescription for
the turbulent component.  For thin, weakly magnetized disks these are
better assumptions than for thick or highly magnetized disks because
the two components do not overlap significantly in the disk. 

The simulations are limited by their duration, which prevents a large
range of radii from reaching quasi-steady state.  They are also
limited by their resolution.  The fastest growing mode of the MRI is
usually not resolved by more than about $10$ grid cells
\citep{penna2010,narayan2012} which is only marginally acceptable
\citep{shiokawa2012,sorathia2012}.  Despite these limitations, these are among the
best GRMHD accretion disk simulations available at present.  Shearing
box simulations can resolve the local physics of the MRI better, but
cannot obtain the dependence of $\alpha$ on radius explicitly.
Nonetheless, the shearing box simulations of \citet{pessah08} should
be revisited with higher resolution (they used $32\times 192\times
32$), as these are the best way to measure the dependence of $\alpha$
on $q$.

Our results suggest that relativistic corrections to $q$ partly
contribute to the higher $\alpha$'s measured in GRMHD simulations
\citep{penna2010,penna2012} compared to Newtonian simulations
\citep{papaloizou2003,fromang2011}.  Assuming $\alpha \propto q^6$ (for $q>0$), we
infer that $\alpha$ is six times larger in the relativistic inner
regions of GRMHD disks than in Newtonian disks.  This
partly resolves the discrepancy but does not go far enough, as the
$\alpha$'s measured from GRMHD simulations are over an order of
magnitude larger than the $\alpha$'s measured from Newtonian
MHD simulations.  More work is needed to understand this discrepancy.  

Switching from a constant $\alpha$ to a radially varying $\alpha(r)$
would have a small effect on black hole spin estimates.
\citet{gou2011} considered the effect on continuum fitting
measurements of the spin of the black hole in Cyg X-1 if one assumes
$\alpha=0.01$ versus $\alpha=0.1$.  The black hole spin decreases
slightly, from $a/M = 0.9988$ to $0.9985$, as $\alpha$ is increased.
Switching from a constant $\alpha$ to the $\alpha(r)$ prescription
will have a similar effect.  This is well below the current
observational sources of error in black hole spin measurements
\citep{kulkarni2011,zhu2012}, so it is not a concern for now.

Black hole spin estimates are restricted to observations of disks
radiating below $30\%$ of the Eddington limit, which corresponds to
thin disks \citep[$h/r\sim 0.1$,][]{kulkarni2011}.  Observations based
on models of thick disks will be more sensitive to the shape of
$\alpha(r)$.

\section*{Acknowledgments}

We thank Eric Blackman for discussions.  This work was supported in
part by NASA grant NNX11AE16G and NSF grant AST-0805832. The
simulations presented in this work were performed in part on the
Pleiades supercomputer, using resources provided by the NASA High-End
Computing (HEC) Program through the NASA Advanced Supercomputing (NAS)
Division at Ames Research Center. We also acknowledge NSF support via
XSEDE resources at NICS Kraken and LoneStar.

\bibliographystyle{mnras}
\bibliography{ms}

\end{document}